\documentclass[twocolumn, superscriptaddress, showpacs,
 amsmath,amssymb,
 aps,
 prx
]{revtex4-2}
\usepackage{graphicx} 
\usepackage{xcolor}

\newcommand{\m}{\text{m}}

\newcommand{\mev}{\text{MeV}}
\newcommand{\gev}{\text{GeV}}

\newcommand{\cm}{\text{cm}}

\newcommand{\mum}{\text{$\mu$}\m}

\newcommand{\mm}{\text{mm}}

\newcommand{\pc}{\text{pC}}
\newcommand{\nc}{\text{nC}}

\newcommand{\mrad}{\text{mrad}}

\newcommand{\hz}{\text{Hz}}
\newcommand{\khz}{\text{kHz}}

\begin{document}

\preprint{APS/123-QED}


\title{Energy-resolved measurement of individual GeV muon tracks generated by electrons from a compact Laser-Plasma Accelerator
}

\author{Davide Terzani}
\email{dterzani@lbl.gov}
\affiliation{%
 Lawrence Berkeley National Laboratory, Berkeley, CA 94720, USA
}
\author{Luc Le Pottier}
\affiliation{%
 Lawrence Berkeley National Laboratory, Berkeley, CA 94720, USA
}
\affiliation{Department of Physics, University of California, Berkeley, CA 94720, USA}
\author{Stanimir Kisyov}
\affiliation{%
 Lawrence Berkeley National Laboratory, Berkeley, CA 94720, USA
}
\author{Pranav Manoj}
\affiliation{%
 Lawrence Berkeley National Laboratory, Berkeley, CA 94720, USA
}
\affiliation{Department of Physics, University of California, Berkeley, CA 94720, USA}
\author{Ryan Heller}
\affiliation{%
 Lawrence Berkeley National Laboratory, Berkeley, CA 94720, USA
}
\author{Maria Mironova}
\affiliation{%
 Lawrence Berkeley National Laboratory, Berkeley, CA 94720, USA
}
\author{Alex Picksley}
\affiliation{%
 Lawrence Berkeley National Laboratory, Berkeley, CA 94720, USA
}
\author{Joshua Stackhouse}
\affiliation{%
 Lawrence Berkeley National Laboratory, Berkeley, CA 94720, USA
}
\affiliation{Department of Nuclear Engineering, University of California, Berkeley, CA 94720, USA}
\author{Hai-En Tsai}
\affiliation{%
 Lawrence Berkeley National Laboratory, Berkeley, CA 94720, USA
}
\author{Raymond Li}
\affiliation{%
 Lawrence Berkeley National Laboratory, Berkeley, CA 94720, USA
}
\affiliation{Department of Nuclear Engineering, University of California, Berkeley, CA 94720, USA}
\author{Timon Heim}
\affiliation{%
 Lawrence Berkeley National Laboratory, Berkeley, CA 94720, USA
}
\author{Maurice Garcia-Sciveres}
\affiliation{%
 Lawrence Berkeley National Laboratory, Berkeley, CA 94720, USA
}
\author{Carlo Benedetti}
\affiliation{%
 Lawrence Berkeley National Laboratory, Berkeley, CA 94720, USA
}
\author{John Valentine}
\affiliation{%
 Lawrence Berkeley National Laboratory, Berkeley, CA 94720, USA
}
\author{Kei Nakamura}
\affiliation{%
 Lawrence Berkeley National Laboratory, Berkeley, CA 94720, USA
}
\author{Anthony J. Gonsalves}
\affiliation{%
 Lawrence Berkeley National Laboratory, Berkeley, CA 94720, USA
}
\author{Jeroen van Tilborg}
\affiliation{%
 Lawrence Berkeley National Laboratory, Berkeley, CA 94720, USA
}
\author{Carl B. Schroeder}
\affiliation{%
 Lawrence Berkeley National Laboratory, Berkeley, CA 94720, USA
}
\affiliation{Department of Nuclear Engineering, University of California, Berkeley, CA 94720, USA}
\author{Eric Esarey}
\affiliation{%
 Lawrence Berkeley National Laboratory, Berkeley, CA 94720, USA
}
\author{Cameron G. R. Geddes}
\affiliation{%
 Lawrence Berkeley National Laboratory, Berkeley, CA 94720, USA
}

\begin{abstract}
Recently, the possibility of LPA-produced muon beams has gained significant interest within the accelerator application community.
Directional, multi-GeV muons can be produced via Bethe-Heitler interactions when multi-GeV electrons hit solid targets.
They are highly penetrating and, thanks to the compactness of the LPA,
offer a path toward a deployable, active muon source.
At the BELLA Center of the Lawrence Berkeley National Laboratory, we previously unambiguously detected muons generated during the interaction of multi-GeV electron beams with a 4 meter-thick electron beam dump.
A new campaign has now extended our diagnostic capabilities to single-muon trajectory reconstruction and energy measurements.
The setup allowed us to individually reconstruct each muon trajectory,
defined by us as a muon passing through three detectors used for the reconstruction.
For a subset of events, we extracted the muon energy from the magnetic-field bending angle, demonstrating production of GeV-scale muons.
This work provides a key demonstration of track-based active-source muography, which enables non-invasive 3D density mapping of concealed or inaccessible samples, and it will accelerate the development of active LPA-based muon sources where compactness, controlled directionality, low divergence, and deep penetration are required.

\end{abstract}


\maketitle

\section{Introduction}

Multi-GeV muons hold great promise as probe particles for transmission radiography and tomography, thanks to their low
scattering cross-section and deep penetration power in matter~\cite{groom_muon_2001}.
In the minimum-ionizing regime, the corresponding mean energy losses are $\simeq 0.5\,\gev/\m$ in rock and concrete,
$\simeq0.2\,\gev/\m$ in water, and $\simeq1.2\,\gev/\m$ in steel.
Cosmic muons have been successfully used to image a magmatic chamber of a volcano~\cite{tanaka_development_2003}, discover a hidden chamber in
Khufu’s Pyramid~\cite{morishima_discovery_2017},
measure the ice-bedrock interface of an alpine glacier~\cite{nishiyama_first_2017},
and image the Unit-1 nuclear reactor of Fukushima Daiichi~\cite{fujii_investigation_2020}.
Studies show that they can be applied in several other contexts~\cite{vanini_muography_2018, procureur_muon_2018}, including underground topology of mines, orebodies, and voids, as well as image structure deficiencies in bridges and critical infrastructure.
While cosmic muons are incident onto the earth's surface at any moment in time,
they are characterized by an average flux at sea level of roughly $1\,\mu/\cm^2\text{min}$,
the majority of which coming at a small angle from the zenith.
Due to the reduced flux, confined directionality, and limited average energy of about $4\,\gev$,
months of exposure are typically necessary to accumulate a significant sensing signal of dense objects.

Traditional proton accelerators are capable of producing large muon fluxes limited to few $\gev$ energies,
however, only a few of these facilities exist and, due to their large scale,
they are not compatible with applications such as on-site muon-based imaging.
Recently, muon production using electron beams generated by compact Laser-Plasma Accelerators (LPAs) has been demonstrated~\cite{terzani_measurement_2025, zhang_proof--principle_2025}.
The accelerating gradients in the plasma can be over 2-3 orders of magnitude larger than in traditional accelerators,
resulting in compact accelerating sections of 10s of cm as opposed to 100s of m.
In such electron-beam-converter schemes, Bethe-Heitler (or lepton pairs) muon production yields $\mathcal{O}(10^4)$ muons per shot
for typical LPA bunch charges of $100\,\pc-1\,\nc$. This corresponds to about 1 muon per $10^5$ incident electrons in the produced muon spectrum.
The spectrum features a high-energy, low-divergence component that is well suited for imaging.

When an electron beam impinges on a high-Z, e.g., lead, tungsten, converter target,
it emits Bremsstrahlung radiation, which consists of gamma rays with energies ranging up to that of the incoming beam confined within a small angle from the beam propagation axis.
When these gamma rays interact with the heavy nuclei of the target,
they can be converted into lepton-antilepton pairs via the Bethe-Heitler process~\cite{chao_handbook_2013}.
For thick converters, while the majority of electrons and positrons typically stop before exiting,
muons and antimuons can propagate through the material only losing a fraction of their energy.
This results in a low-divergence (roughly $\lesssim 100\,\mrad$ for $\simeq 10\,\gev$ initial electron beam energies) muon beam with a Bremsstrahlung-like energy spectrum with maximum energy equal to that of the initial electron beam,
minus the energy lost in the converter.
In addition to pair production, a substantial amount of muons are generated via the decay of mesons produced during the Bremsstrahlung interaction, namely via meson photoproduction~\cite{titov_dimuon_2009}.
This produces a large number of uncollimated, i.e., almost isotropic, low-energy muons,
where the energy is limited by the distance from the source due to the relativistic dilation of the high-energy meson decay time.
A detailed description of the electron beam-based muon production can be found in several references~\cite{titov_dimuon_2009, rao_bright_2018, calvin_laser-driven_2023, terzani_measurement_2025}. 
The Bethe-Heitler muons are particularly well-suited for imaging applications,
where three-dimensional density and material specificity of sample environments
can be derived from measurements of both the incoming and the transmitted, deflected muon tracks.
With an optimized sample/object illumination, enabled by the controlled directionality of an LPA-based muon source,
the exposure time required to collect significant tomography datasets can be reduced by orders of magnitude compared to cosmic-ray muons,
in particular if future laser and LPA-based muon systems enable $\khz$ operation.
In many fields, such as mining, civil engineering, and national security, these capabilities could be transformative, because no other approach currently enables probing large, dense, deep samples in timescales compatible with operational needs.

Recent experiments have investigated LPA-generated muons~\cite{terzani_measurement_2025, zhang_proof--principle_2025, calvin_experimental_2026, dreesen_detection_2014}.
In these studies, the main observable was the presence of muons within the electromagnetic shower following a thick target.
However, separating the muon signal from the substantial background of secondary radiation remained challenging.
As a result, muons identification relied on statistical averaging over many laser shots.
Consequently, these studies did not enable shot-by-shot proof of the muon yield,
nor did they perform any single-particle tracking.
At the BELLA Center, we performed experiments using a novel diagnostic to measure the energy and propagation direction of the high-energy component of the muon spectrum.
In this paper, we report for the first time single track reconstruction
and energy measurement of multi-GeV LPA-generated muons. 
We developed a muon telescope consisting of two stacks of three silicon-based particle trackers, with a dipole magnet placed between the stacks.
Each stack identifies muons by their straight trajectory, as opposed to other background particles that scatter or get absorbed during the propagation,
and the dipole imposes a well-defined angular kick to the muons, enabling single-muon energy measurements.
This diagnostic technique is particularly well suited for detecting muons generated by an LPA.
Silicon-based trackers are relatively insensitive to the associated photon background, provided the photon flux remains sufficiently low.
This combination enables high-precision track reconstruction,
and, when a sample is placed in the muon path, yields information on muon energy and scattering angles. 
The results highlight the applicability of the multi-stack muon detectors to the LPA muon source, and thus showcase the powerful potential of LPA-based, multi-GeV muon tomographic applications.
Note that while the current repetition rate of these lasers and accelerators is limited to $\lesssim 1\,\hz$,
novel near-term LPA-driving laser development is focusing on $\khz$-scale operation~\cite{kiani_high_2023}, which will enhance applicability for deployable, high-repetition-rate muography. 
The first track-resolved LPA muon results presented in this paper provide key demonstrations towards track-based active-source muography, ideal for applications where scattering-angle retrieval and/or single-muon energy information will significantly boost muographic imaging resolution.
This paper is organized as follows.
In Section~\ref{sec:exp_setup} we discuss the experimental setup, including the custom-built muon telescope.
in Section~\ref{sec:exp_results} we present the experimental results, showing muon tracking and energy reconstruction. 
Finally, in Section~\ref{sec:conclusions} we discuss the conclusions.

\section{Experimental Setup}\label{sec:exp_setup}

In this Section, we will present the experimental setup used to generate and detect a high-energy, directional muon beam.
The LPA and converter target setups are similar to what was reported in our previous work~\cite{picksley_matched_2024, terzani_measurement_2025},
with the difference that a longer, $40\,\cm$-long gas-jet was used as a plasma target.

\begin{figure*}[!ht]
\includegraphics[width=0.98\linewidth]{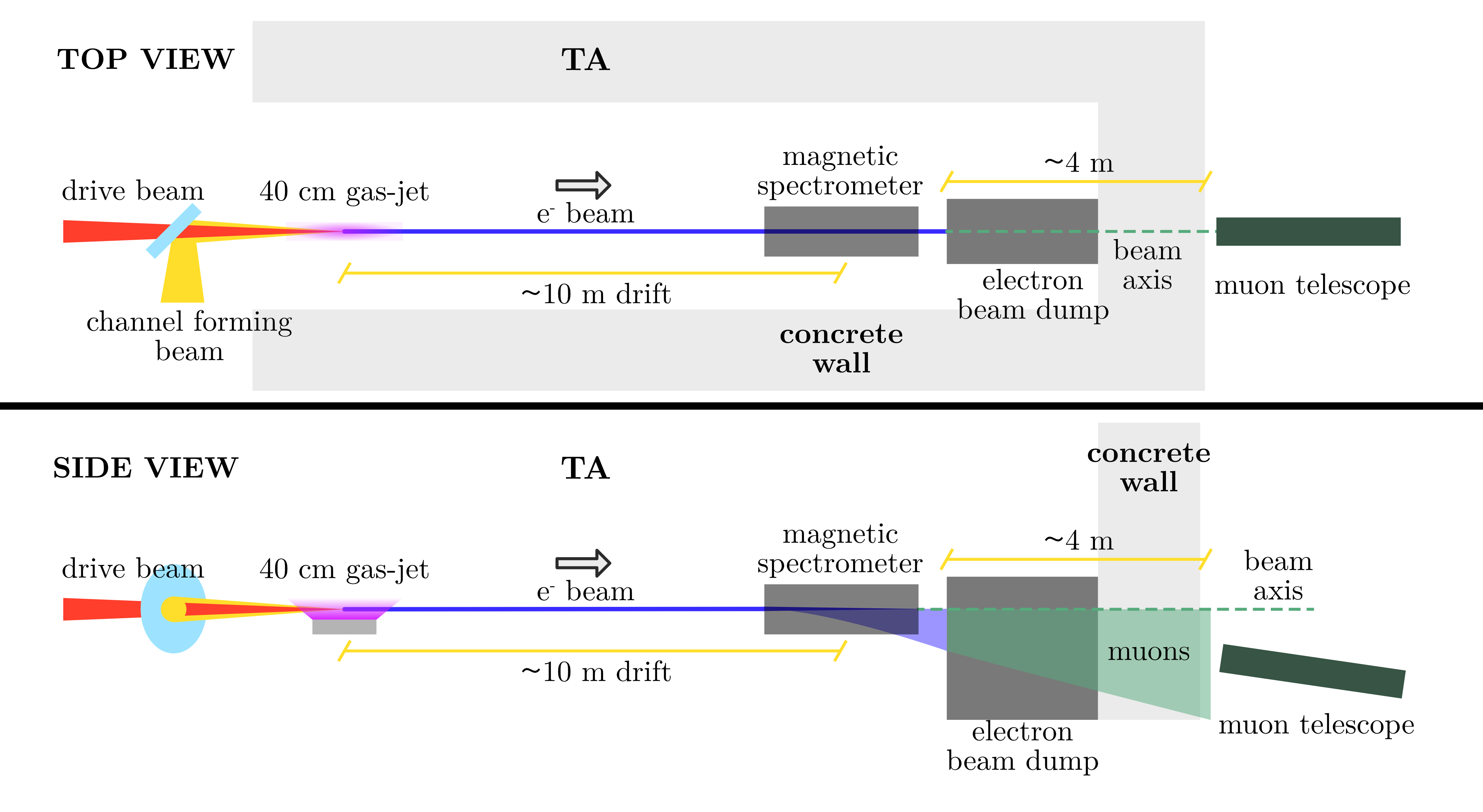}
\caption{Schematic of the apparatus used to generate and detect muons.
A channel forming beam~\cite{shalloo_hydrodynamic_2018, shalloo_low-density_2019, smartsev_axiparabola_2019, morozov_ionization_2018, picksley_meter-scale_2020, feder_self-waveguiding_2020},  and a drive laser beam power the LPA, which produces electron beams up to $10\,\gev$.
These beams pass through a dipole (magnetic spectrometer) which disperses them on the vertical axis and are then stopped in the high-Z layers of the electron beam dump.
Here, they generate muons in a narrow cone that propagate through the concrete that encloses the Target Area (TA) and reach a muon telescope, placed directly behind the wall along the beam propagation axis.}
\label{fig:apparatus}
\end{figure*}

\subsection{LPA-based muon source}

\begin{figure}[!ht]
\includegraphics[width=0.98\linewidth]{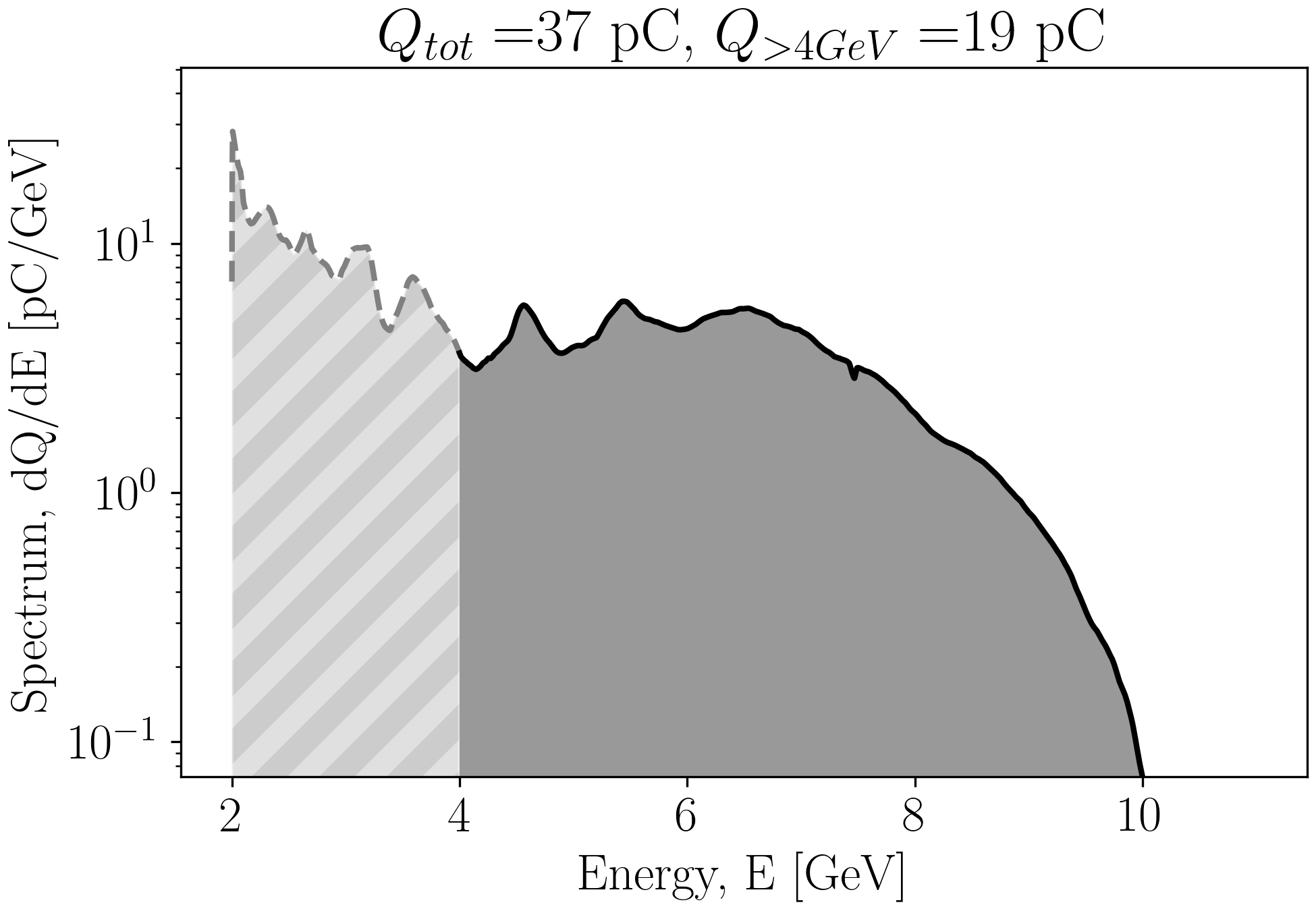}
\caption{Average electron beam spectrum derived from all shots that generated muons.
To generate a measurable muon event, electron beams must contain charge at energies above $4\,\gev$,
which corresponds to the minimum energy dissipated while traversing the shielding to reach the detector.
The hatched region indicates the portion of the spectrum that does not contribute to the generation of a measurable muon flux in our setup.
During this run, we produced electron beams with energies up to $10\,\gev$.}
\label{fig:average_e_beam}
\end{figure}

The source for the muon beam was the LPA-driven electron beam generated using the BELLA PW laser~\cite{nakamura_diagnostics_2017}.
Details about the accelerator can be found in Refs.~\cite{picksley_matched_2024, terzani_measurement_2025} and a schematic of the experiment is presented in Figure~\ref{fig:apparatus}.
With this system, we generated electron beams with energy up to $10\,\gev$ at $0.1\,\hz$ for a total of $\sim20$ hours across 3 days of operation, see Figure~\ref{fig:average_e_beam}.

The high-energy electrons produced in the LPA are passed through a 1.08 T dipole (labeled ``magnetic spectrometer" in Figure~\ref{fig:apparatus}) which bends the electron beams towards the ground.
Then, the electrons are stopped in a multi-layer dump composed of a $40.5\,\cm$ layer of lead, a $1\,\m$ layer of steel,
and a final $1.8\,\m$ layer of concrete.
The dump is attached to an additional $90\,\cm$ concrete wall, which delimits the Target Area (TA) where the beams are generated.
Electron beams passing through the shielding are stopped and generate muons, among other secondaries,
via the previously described mechanisms.
High-energy muons generated in the bulk of the beam dump are able to traverse the remaining shielding and
escape through its rear into the adjacent room, where we placed the detectors.
In order to favor measurements of directional, high-energy muons,
we placed the detector directly behind the wall and aligned with the electron beam propagation axis, as shown in Figure~\ref{fig:apparatus}.

\subsection{Detectors}

\begin{figure}[htbp]
\includegraphics[width=0.98\linewidth]{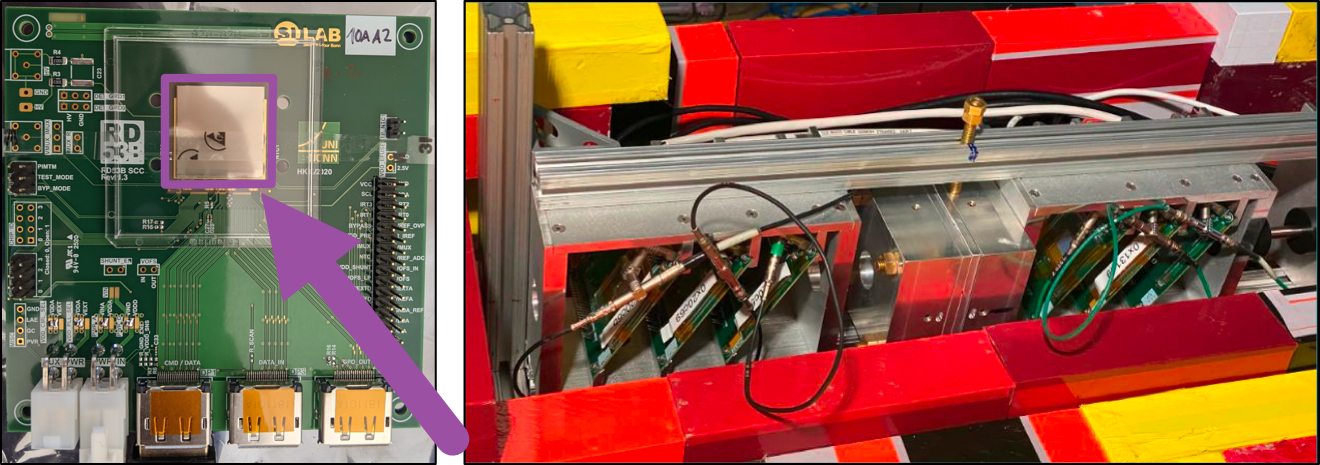}
\includegraphics[width=0.98\linewidth]{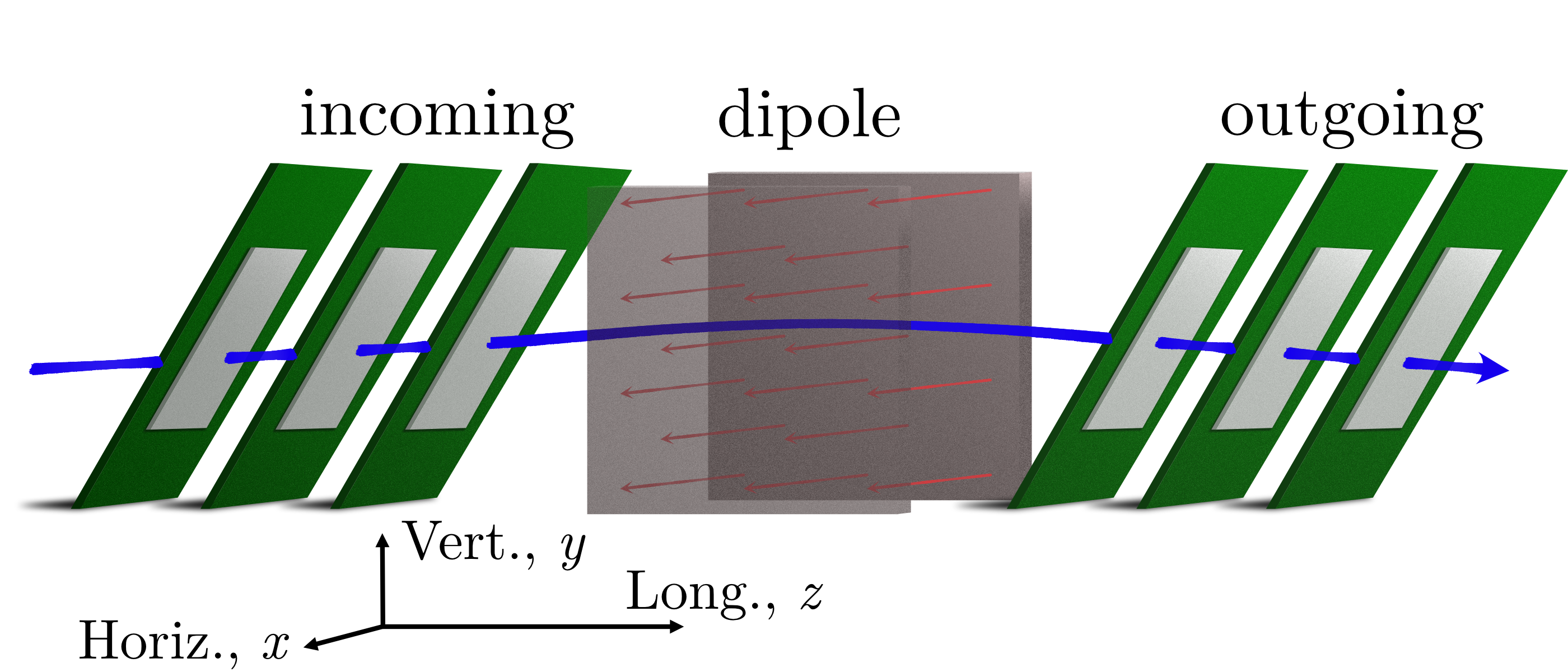}
\caption{(Top) Picture of the muon telescope with the two 3-plane detector module stacks on each side of the Hallbach magnet array
inside of a lead brick housing. The left panel shows a picture of one of the chips (highlighted in purple).
(Bottom) Schematic of the muon telescope.
The electron beam is coming from the left with respect to this picture.
We call the left triplet incoming stack, before the magnet, and the right triplet outgoing stack.
The incoming muons propagate along the longitudinal axis $z$.
The magnetic field is oriented along the horizontal axis $x$,
and particles are bent along the vertical axis $y$.}
\label{fig:telescope_picture}
\end{figure}

The muon telescope consists of two stacks of three silicon tracking detectors,
with an intervening Halbach magnet in between.
Pictures and schematic of the telescope are shown in Figure~\ref{fig:telescope_picture}.
Each tracking detector module consists of a single ATLAS ITkPix readout ASIC~\cite{alimonti_rd53_2025}
bump bonded to a planar silicon sensor.
The ITkPix ASIC is a 65~nm feature size CMOS technology chip with a 50~$\mu$m by 50~$\mu$m pixel size
and a 384$\times$400 pixel matrix size, providing a total active chip area of $1.92\times2\,\cm^2$.
The sensors for all six tracking detectors are n-in-p planar sensors with a sensor thickness of either 100~$\mu$m or 150~$\mu$m,
operated above their respective depletion voltages of 15/35 V.
Both sensor topologies have been shown to have nearly 100\% charged particle detection efficiency after production~\cite{samy_recent_2025}. 
Each detector module is mounted to a Print Circuit Board (PCB), referred to as a Single Chip Card (SCC) and
shown in Figure~\ref{fig:telescope_picture} (top left panel), to allow powering and readout.
Three such modules are housed within an extruded aluminum support structure to form a single detector stack.
Inside this structure, the detector modules are positioned at a 30$^\circ$ angle with respect to the beam axis.
Consequently, muons traversing the chip deposit charge across multiple pixels.
Because the ITkPix readout ASIC is capable of digitizing the measured charge for each pixel hit, an analog geocenter
can be calculated.
This improves the spatial resolution of the detector module below the $50\,\mum$ pixel quantization limit to
$50~\mu\text{m}/\sqrt{12} = 14.4~\mu\text{m}$
in the direction the detector is module rotated in.
Finally, to improve the momentum resolution, the modules are rotated in the same direction as the magnet field kick applied between detector stacks. 

The ITkPix readout ASIC timestamps events in 25~ns intervals
and events are read out upon reception of a trigger signal from the data acquisition.
The muon telescope is triggered if coincident hits are registered in the first and third plane of either stack.
Such a trigger choice enables recording not only muons traversing the full telescope,
but also muons within the acceptance of only one of the two stacks.
The momentum of muons traversing only one stack cannot be measured,
but they can nevertheless be used for alignment purposes within the stack.
This trigger scheme is proven to be very resilient against electron and photon background,
that do not result in coincident hits in multiple planes.

\subsection{Alignment strategy and detector calibration}
The detector planes within each stack are separated by $40\,\mm$, and the two stacks are separated by $220\,\mm$.
A Halbach magnet array is centered between the two stacks.
The field of the Halbach magnet extends for an effective length of $151\,\mm$ along the beam axis.
The average field over this length is $B\simeq 0.29\,\text{T}$.
Due to the limited statistics available in the LPA runs,
we used cosmic-ray muons to establish the initial alignment of the telescope.
To collect sufficient statistics, the full telescope was oriented vertically and kept in this configuration for an extended period (on the order of 1 month continuous operation).
These data were then used to extract the alignment parameters, which were applied to the subsequent analysis of the LPA runs.
In particular, they provide a reference to calibrate the relative positions of the three detector modules within each stack.
While the internal alignment of each stack is highly stable thanks to the rigid aluminum frame,
due to the telescope length, the relative alignment between the two stacks is more sensitive to changes in orientation:
after rotating the telescope from vertical to horizontal prior to the LPA measurements, a small angular offset between the stacks can be introduced.
This inter-stack angular offset is determined from the tracks of LPA-generated muons measured in the horizontal configuration.

\subsection{Muon track reconstruction}

Electron beams traversing the high-Z layers of the electron beam dump emit Bremsstrahlung radiation that is converted into muon pairs.
By placing the detector directly behind the dump and aligning it with the beam propagation axis,
we maximize the chances of detecting these muons,
which are generated within a narrow cone around the axis.
Other muon sources, such as those resulting from pion decay, exhibit a much broader angular distribution and lower energies.
These are stopped by the strong filtering power of the dump, resulting in negligible background contamination~\cite{terzani_measurement_2025}.

To be unequivocally identified, a muon must traverse all three detectors in a triplet.
We discarded tracks defined by only two points in our analysis.
This requirement enables the identification of muonic hits and distinguishes them from background hits caused by photons or other charged particles.
The procedure is only effective in a low-occupancy regime, i.e., when few detector pixels are activated per event.
We discussed the background radiation expected in our LPA setup in a previous work~\cite{terzani_measurement_2025}
where we identified a significant photon flux and some neutrons behind the beam dump.
Although individual photon interaction probability is low, a high flux could obscure clean measurements.
To mitigate this, we installed a 10 cm-thick lead shield around the muon telescope.
Tracks are extracted from the detector hits using Corryvreckan~\cite{dannheim_corryvreckan_2021}.
The software calibrates the detector relative distances and angles
and performs fits to extract the trajectory parameters (geocenters) for each detector.
This enables estimating the incoming and outgoing slopes and calculating the bending angle in case of a muon passing through both stacks.
The software automatically rejects spurious hits via a combinatorics procedure,
where straight lines are fitted through all three detectors.
With multiple spurious hits, trajectory reconstruction fails, and muons cannot be tracked.
Optimal detector positioning and custom shielding reduced the spurious hits generated by photons and neutrons and provided tracking information for the detected muons.

\section{Experimental Results}\label{sec:exp_results}

Muon detection was performed in parallel with our laser-plasma acceleration campaign,
whose primary goal was the optimization of $10\,\gev$ electron-beam production.
To map the relevant LPA parameter space, we repeatedly modified the LPA configuration, which resulted in intermittent (discontinuous) electron beams.

For a muon to be detectable in our experimental setup,
the incident electron beam must contain a non-negligible population of electrons above approximately $4\,\gev$.
Numerical analysis presented in our previous work~\cite{terzani_measurement_2025}
indicated that a muon traversing the metal and concrete shielding of the experimental hall loses approximately $4\,\gev$.
Electrons with energies below this threshold therefore do not generate muons measurable with our setup.
From the pool of discontinuous electron-beam shots, we define as electron-beam candidates
those shots in which the charge in the energy interval above $4\,\gev$ is $\geq 5\,\pc$.
Across a total of 20 hours of operation, we recorded 361 electron beam candidates.
These electron beams exhibited high maximum energies, up to $10\,\gev$,
together with a significant charge content above $4\,\gev$.
Figure~\ref{fig:average_e_beam} shows the average spectrum collected
from all the electron beams that resulted in detected muon events.
The gray area represents the spectral region that does not contribute to measurable muon generation.
For this spectrum, the total charge is $Q_{tot}=37\,\pc$
and the muon-producing charge is $Q_{>4\gev}=19\,\pc$.

While the stability of an LPA system is of primary importance for achieving application-ready status,
additional work is required to ensure the consistent generation of stable, high-energy electron beams at every laser shot.
Since this optimization is beyond the scope of the present work,
we quantify the efficiency of our source solely in terms of
the number of electron beams effectively able to generate measurable muons.

From the 361 incoming electron beams, we recorded a total of 39 muon trajectories crossing either stack.
Of these, 10  were full-stack events, where full-stack indicates events with a muon crossing both detector stacks.
With the telescope positioned horizontally, the contribution of cosmic events is negligible.
\begin{figure}[!ht]
\includegraphics[width=0.98\linewidth]{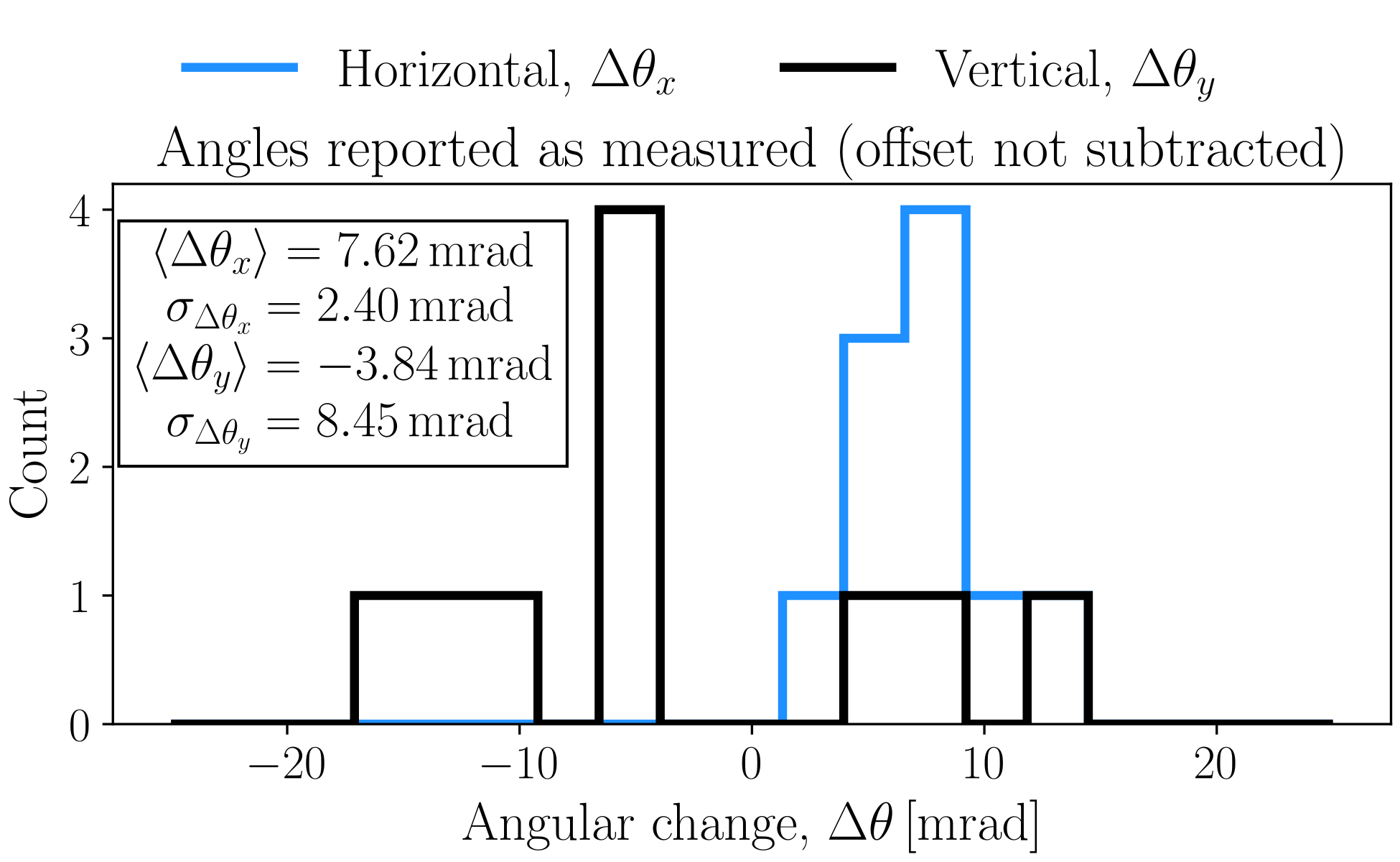}
\caption{Histogram of the angular deflection $\Delta \theta=\theta_{\mathrm{out}}-\theta_{\mathrm{inc}}$ for both the horizontal (blue line)
and vertical (black line) axes.
The effect of the magnetic field spreading is clear along the vertical axis.
The histograms show that there is (on average) a constant angular offset between the two stacks.}
\label{fig:angular_offset}
\end{figure}

The muon momentum $p$ is computed via the magnetic rigidity as
\begin{equation}
    B\rho\,[\text{T}\cdot\m] = 3.36\,pc\,[\gev],
\end{equation}
which, accounting for the effective field length $151\,\mm$, it can be rewritten as
\begin{equation}
    pc\,[\gev]\simeq 4.5\times 10^{-2} B[\text{T}]/\sin(\alpha),
\end{equation}
where $\alpha=\theta_{\mathrm{out}}-\theta_{\mathrm{inc}}-\theta_{\mathrm{offset}}$ is the deflection angle.
The muon kinetic energy can be calculated from the momentum via
\begin{equation}
    E = \sqrt{m_{\mu}^2c^4+p^2c^2}-m_{\mu}c^2,
\end{equation}
where $m_{\mu}c^2=105.7\,\mev$ is the mass of the muon.
The error on the energy estimation is dominated by the uncertainty on the trajectory slope reconstruction and on the angular offset between the two detector stacks.
The former can be estimated via accumulating a large amount of straight cosmic-muon trajectories.
The residuals from the track reconstruction show a shot-to-shot uncertainty of $\sigma_{\theta}=2.5\,\mrad.$
The source of this uncertainty requires further investigation.
Cosmic muons cannot be used to estimate the angular offset between the two stacks,
as the telescope is very sensitive to changes in orientations.
We used the 10 full-stack muon tracks from the LPA to estimate this offset under the assumption that the angular distribution
is expected to be symmetric along both the horizontal and vertical axes.
Figure~\ref{fig:angular_offset} shows the collected data from full-stack,
LPA-generated muonic events and we can see that there is an average angular offset between the two stacks. 
The offset on the vertical (bending) axis is $\theta_{\mathrm{offset}}=\left(-3.84 \pm 8.45/\sqrt{10}\right)\,\mrad$.
The large uncertainty is due to limited statistics from the LPA.
Adapting the telescope assembly technique could provide greater robustness against movements
and would enable us to use cosmic rays to obtain a more precise estimate of the angular offset between the two stacks.
Longer exposure to LPA muons, a higher detector acceptance (for instance using larger silicon chips),
and future high-repetition rate laser systems increase the statistical significance of the muon data from the LPA,
leading to a faster and more precise detector calibration.

\begin{figure*}[!ht]
\includegraphics[width=0.7\linewidth]{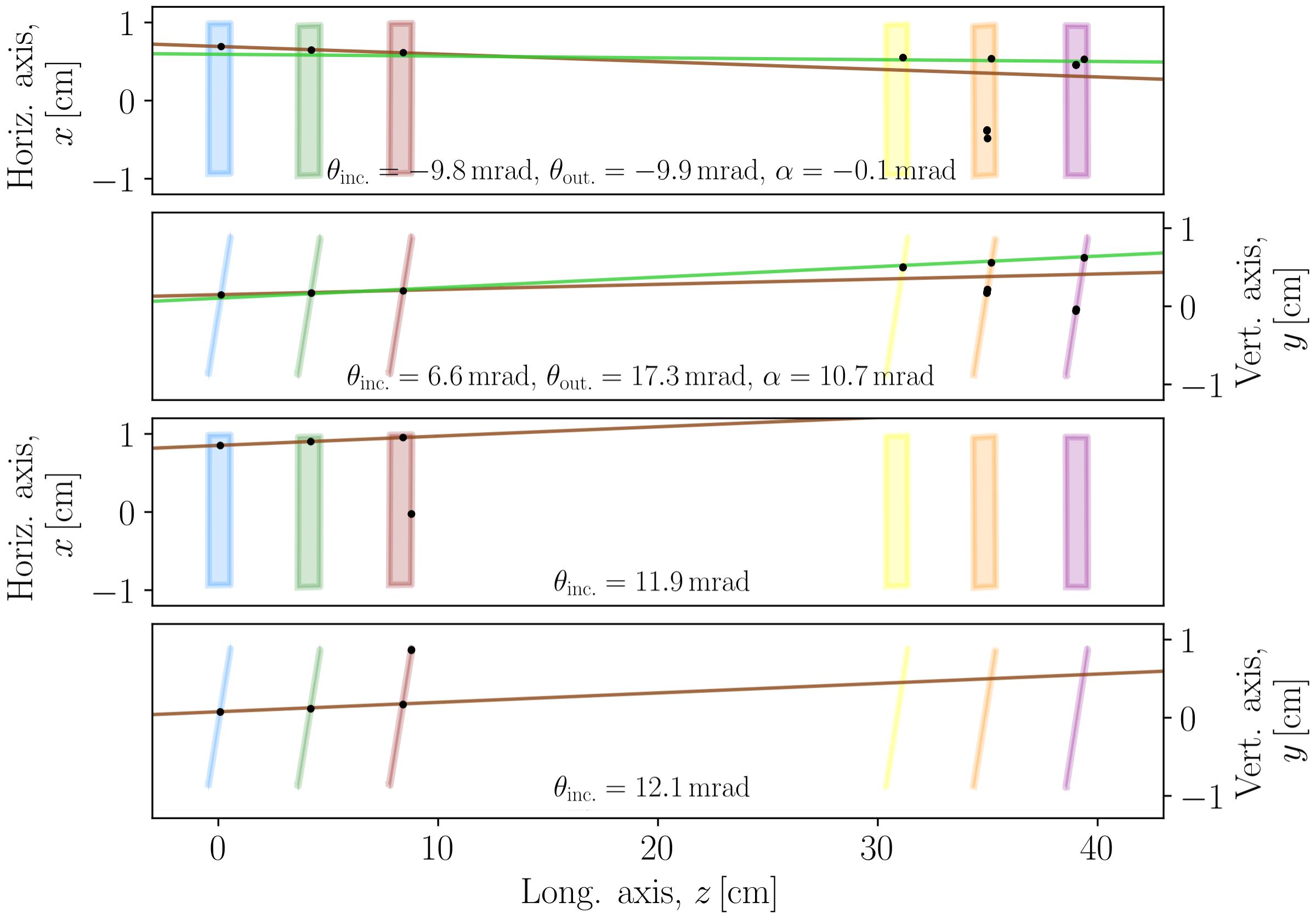}
\caption{Two examples of muon tracks, one passing through both stacks the other passing through only one.
The top two panels refer to the full-stack event (top-view and side-view, respectively).
Trajectories are reconstructed using Corrywreckan to fit straight lines through the detector triplets.
The brown line shows the incoming trajectory, the green line the outgoing one.
The magnet bends the trajectory along the vertical axis.
The different slopes between incoming and outgoing trajectories on the horizontal axis is due to the angular offset between the two stacks.
The bottom two panels show the analogous data for a muonic event that only passes through the incoming stack.
No momentum estimate is possible in this case.
Every black dot on the detectors is a registered hit.
Each detector is identified with a different color, in sequence.
The magnetic field is oriented along the $x$ axis,
therefore particles are dispersed in the $y-z$ plane.
Note that because the figures use an aspect ratio different from 1,
the angles displayed in the images are not the real angles.}
\label{fig:track_example}
\end{figure*}

Figure~\ref{fig:track_example} shows the track reconstruction of two muonic events.
In the top two panels, we show the top and side views, respectively, of a full-stack event.
The vertical axis ($y$) is the bending axis for the magnetic field of the dipole.
The muon shown has an estimated kinetic energy $0.4\lesssim E \lesssim 1.0\,\gev$.
The bottom two panels show a single-stack muonic event: the muon crosses only the incoming stack, missing the outgoing one. We cannot estimate the energy of such a muon.
A few background hits are visible in the pictures.
These hits can be identified by their small cluster size
(photonic hits generate very localized signals)
and by the fact that they do not align along a straight line.
Data demonstrate extremely clean measurements in which muons are easily identifiable,
as shown in Figures~\ref{fig:track_example}.
Note that the signals shown are raw; no background subtraction is necessary to identify the muon trajectories.
Achieving this is particularly challenging, given that LPA beams produce a short burst of secondary radiation capable of blinding a detector.

\begin{figure}[!ht]
\includegraphics[width=0.98\linewidth]{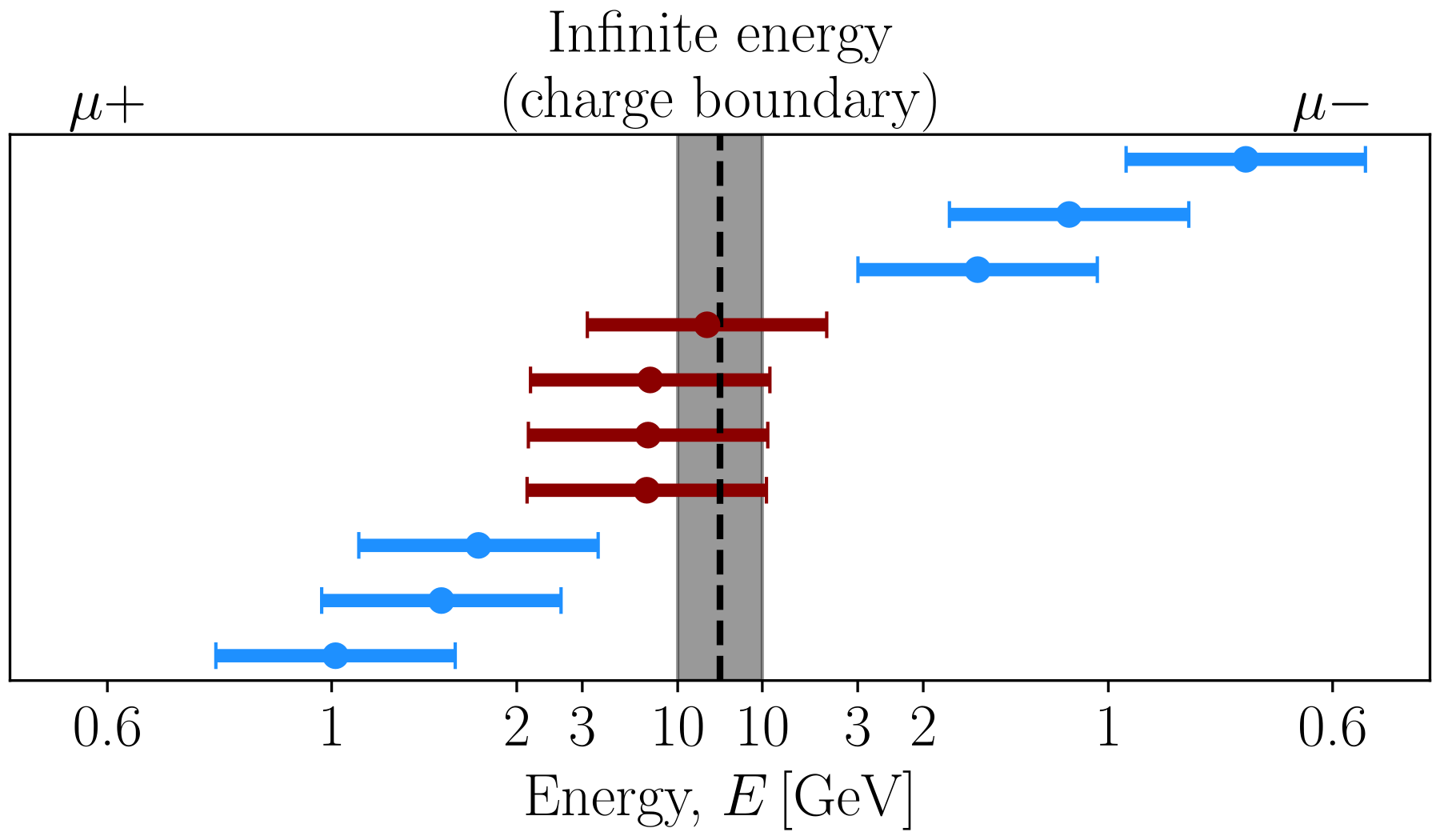}
\caption{Muon energies and uncertainties for the 10 full-stack events produced by the LPA.
The energy is parametrized by $1/\sin(\alpha)$.
The energy range of each event corresponds to angular variation of $\pm\sigma_{\alpha}$ with respect to the central value.
The left side of the panel (negative bending) corresponds to positive muons,
the right side of the panel (positive bending) to negative muons.
In red we show tracks with ambiguous charge due to the angular uncertainty.
The central gray band marks the muon-energy range that cannot be produced because of the limited initial electron-beam energy.
All these tracks show GeV-level muon energies, in agreement with the expected muon production from $10\,\gev$ electron beams,
which also accounts for $4\,\gev$ lost in traversing the shielding.}
\label{fig:energy}
\end{figure}
The muon energies associated with the 10 full-stack events and their uncertainties are shown in Figure~\ref{fig:energy}.
The energy axis is parametrized by $1/\sin(\alpha)$.
The left side of the panel shows negative bending angles (corresponding in our configuration to positive muons),
while the right side corresponds to negative muons.
The energy range for each muon is shown to account for angular deviations of $\pm\sigma_{\alpha}$ in both directions with respect to the central value.
For some muon events, marked in red, the bending angle is compatible with both directions and therefore they have an ambiguous charge sign.
Nonetheless, it is possible to determine the minimum energy associated with the muon.
The measurements show that, within one $\sigma_\alpha$, all muons had energies $E\gtrsim 1\,\gev$,
which is compatible with the expectations for our setup, given electron-beam energies up to $10\,\gev$ and an estimated $4\,\gev$ energy loss in traversing the shielding.

A primary limiting factor in the current measurement was the detector acceptance.
With an angular aperture of $\sim6\times 10^{-6}\,\text{sr}$ and a detector surface of $1.92\times2\,\cm^2$,
the setup captured only a small fraction of the generated muons.
Our previous studies indicated an expected yield on the order of $100$ directional muons per incoming electron beam,
spread over a surface area of approximately $1\,\m^2$,
which is significantly larger than the active region covered by our current detectors.
Nevertheless, this work establishes a critical foundation for detailed source characterization, which is essential for understanding how to optimize and tune LPA-driven muon sources for future applications.

Overall, this campaign demonstrates the successful detection of muon tracks produced in LPA interactions and the corresponding energy reconstruction,
which prove the generation of muons with GeV-scale energies.
Despite the intrinsic experimental difficulties associated with burst-like radiation and limited statistics, the muon trajectories reconstructed in the telescope are clean and unambiguous.

\section{Conclusions}\label{sec:conclusions}

We have presented a first-time measurement of muon track reconstruction and energy estimation using an LPA-based source.
By using a telescope comprising two stacks of silicon trackers with an intervening magnet, we identified a total of 39 muon tracks from 361 incoming electron beam candidates.
We performed a direct energy measurement on 10 of these muon tracks,
demonstrating the production of muons with energies $E\gtrsim 1\,\gev$ within one $\sigma_\alpha$.
This is consistent with the expected values from a $10\,\gev$ electron beam and with our previous simulation studies.

Our results provide a key demonstration for track-based active-source muography, which is critical for applications where scattering-angle retrieval and/or single-muon energy information will significantly boost muographic imaging resolution at short exposure times.
Finally, because the muon yield scales linearly with the laser repetition rate, moving from Hz-scale operation to kHz operation would increase the muon arrival rate by orders of magnitude, enabling fast image reconstructions at short exposure times.
For certain applications, this capability is transformative because it offers a
route to interrogate the interior of samples that are otherwise inaccessible to non-penetrative measurements.

\begin{acknowledgments}

\noindent This work was supported by the Director, Office of Science, Office of High Energy Physics, of the U.S. Department of Energy under Contract No. DE-AC02-05CH11231, the Defense Advanced Research Projects Agency (DARPA), and used the computational facilities at the National Energy Research Scientific Computing Center (NERSC) under award HEP-ERCAP0035612.
\end{acknowledgments}

\bibliography{BellaMu}

\begin{thebibliography}{27}%
\makeatletter
\providecommand \@ifxundefined [1]{%
 \@ifx{#1\undefined}
}%
\providecommand \@ifnum [1]{%
 \ifnum #1\expandafter \@firstoftwo
 \else \expandafter \@secondoftwo
 \fi
}%
\providecommand \@ifx [1]{%
 \ifx #1\expandafter \@firstoftwo
 \else \expandafter \@secondoftwo
 \fi
}%
\providecommand \natexlab [1]{#1}%
\providecommand \enquote  [1]{``#1''}%
\providecommand \bibnamefont  [1]{#1}%
\providecommand \bibfnamefont [1]{#1}%
\providecommand \citenamefont [1]{#1}%
\providecommand \href@noop [0]{\@secondoftwo}%
\providecommand \href [0]{\begingroup \@sanitize@url \@href}%
\providecommand \@href[1]{\@@startlink{#1}\@@href}%
\providecommand \@@href[1]{\endgroup#1\@@endlink}%
\providecommand \@sanitize@url [0]{\catcode `\\12\catcode `\$12\catcode `\&12\catcode `\#12\catcode `\^12\catcode `\_12\catcode `\%12\relax}%
\providecommand \@@startlink[1]{}%
\providecommand \@@endlink[0]{}%
\providecommand \url  [0]{\begingroup\@sanitize@url \@url }%
\providecommand \@url [1]{\endgroup\@href {#1}{\urlprefix }}%
\providecommand \urlprefix  [0]{URL }%
\providecommand \Eprint [0]{\href }%
\providecommand \doibase [0]{https://doi.org/}%
\providecommand \selectlanguage [0]{\@gobble}%
\providecommand \bibinfo  [0]{\@secondoftwo}%
\providecommand \bibfield  [0]{\@secondoftwo}%
\providecommand \translation [1]{[#1]}%
\providecommand \BibitemOpen [0]{}%
\providecommand \bibitemStop [0]{}%
\providecommand \bibitemNoStop [0]{.\EOS\space}%
\providecommand \EOS [0]{\spacefactor3000\relax}%
\providecommand \BibitemShut  [1]{\csname bibitem#1\endcsname}%
\let\auto@bib@innerbib\@empty
\bibitem [{\citenamefont {Groom}\ \emph {et~al.}(2001)\citenamefont {Groom}, \citenamefont {Mokhov},\ and\ \citenamefont {Striganov}}]{groom_muon_2001}%
  \BibitemOpen
  \bibfield  {author} {\bibinfo {author} {\bibfnamefont {D.~E.}\ \bibnamefont {Groom}}, \bibinfo {author} {\bibfnamefont {N.~V.}\ \bibnamefont {Mokhov}},\ and\ \bibinfo {author} {\bibfnamefont {S.~I.}\ \bibnamefont {Striganov}},\ }\bibfield  {title} {\bibinfo {title} {Muon stopping power and range tables 10 {MeV}–100 {TeV}},\ }\href {https://doi.org/10.1006/adnd.2001.0861} {\bibfield  {journal} {\bibinfo  {journal} {Atomic Data and Nuclear Data Tables}\ }\textbf {\bibinfo {volume} {78}},\ \bibinfo {pages} {183} (\bibinfo {year} {2001})}\BibitemShut {NoStop}%
\bibitem [{\citenamefont {Tanaka}\ \emph {et~al.}(2003)\citenamefont {Tanaka}, \citenamefont {Nagamine}, \citenamefont {Kawamura}, \citenamefont {Nakamura}, \citenamefont {Ishida},\ and\ \citenamefont {Shimomura}}]{tanaka_development_2003}%
  \BibitemOpen
  \bibfield  {author} {\bibinfo {author} {\bibfnamefont {H.}~\bibnamefont {Tanaka}}, \bibinfo {author} {\bibfnamefont {K.}~\bibnamefont {Nagamine}}, \bibinfo {author} {\bibfnamefont {N.}~\bibnamefont {Kawamura}}, \bibinfo {author} {\bibfnamefont {S.~N.}\ \bibnamefont {Nakamura}}, \bibinfo {author} {\bibfnamefont {K.}~\bibnamefont {Ishida}},\ and\ \bibinfo {author} {\bibfnamefont {K.}~\bibnamefont {Shimomura}},\ }\bibfield  {title} {\bibinfo {title} {Development of a two-fold segmented detection system for near horizontally cosmic-ray muons to probe the internal structure of a volcano},\ }\href {https://doi.org/https://doi.org/10.1016/S0168-9002(03)01372-X} {\bibfield  {journal} {\bibinfo  {journal} {Nuclear Instruments and Methods in Physics Research Section A: Accelerators, Spectrometers, Detectors and Associated Equipment}\ }\textbf {\bibinfo {volume} {507}},\ \bibinfo {pages} {657} (\bibinfo {year} {2003})}\BibitemShut {NoStop}%
\bibitem [{\citenamefont {Morishima}\ \emph {et~al.}(2017)\citenamefont {Morishima}, \citenamefont {Kuno}, \citenamefont {Nishio}, \citenamefont {Kitagawa}, \citenamefont {Manabe}, \citenamefont {Moto}, \citenamefont {Takasaki}, \citenamefont {Fujii}, \citenamefont {Satoh}, \citenamefont {Kodama}, \citenamefont {Hayashi}, \citenamefont {Odaka}, \citenamefont {Procureur}, \citenamefont {Attié}, \citenamefont {Bouteille}, \citenamefont {Calvet}, \citenamefont {Filosa}, \citenamefont {Magnier}, \citenamefont {Mandjavidze}, \citenamefont {Riallot}, \citenamefont {Marini}, \citenamefont {Gable}, \citenamefont {Date}, \citenamefont {Sugiura}, \citenamefont {Elshayeb}, \citenamefont {Elnady}, \citenamefont {Ezzy}, \citenamefont {Guerriero}, \citenamefont {Steiger}, \citenamefont {Serikoff}, \citenamefont {Mouret}, \citenamefont {Charlès}, \citenamefont {Helal},\ and\ \citenamefont {Tayoubi}}]{morishima_discovery_2017}%
  \BibitemOpen
  \bibfield  {author} {\bibinfo {author} {\bibfnamefont {K.}~\bibnamefont {Morishima}}, \bibinfo {author} {\bibfnamefont {M.}~\bibnamefont {Kuno}}, \bibinfo {author} {\bibfnamefont {A.}~\bibnamefont {Nishio}}, \bibinfo {author} {\bibfnamefont {N.}~\bibnamefont {Kitagawa}}, \bibinfo {author} {\bibfnamefont {Y.}~\bibnamefont {Manabe}}, \bibinfo {author} {\bibfnamefont {M.}~\bibnamefont {Moto}}, \bibinfo {author} {\bibfnamefont {F.}~\bibnamefont {Takasaki}}, \bibinfo {author} {\bibfnamefont {H.}~\bibnamefont {Fujii}}, \bibinfo {author} {\bibfnamefont {K.}~\bibnamefont {Satoh}}, \bibinfo {author} {\bibfnamefont {H.}~\bibnamefont {Kodama}}, \bibinfo {author} {\bibfnamefont {K.}~\bibnamefont {Hayashi}}, \bibinfo {author} {\bibfnamefont {S.}~\bibnamefont {Odaka}}, \bibinfo {author} {\bibfnamefont {S.}~\bibnamefont {Procureur}}, \bibinfo {author} {\bibfnamefont {D.}~\bibnamefont {Attié}}, \bibinfo {author} {\bibfnamefont {S.}~\bibnamefont {Bouteille}}, \bibinfo {author} {\bibfnamefont {D.}~\bibnamefont {Calvet}},
  \bibinfo {author} {\bibfnamefont {C.}~\bibnamefont {Filosa}}, \bibinfo {author} {\bibfnamefont {P.}~\bibnamefont {Magnier}}, \bibinfo {author} {\bibfnamefont {I.}~\bibnamefont {Mandjavidze}}, \bibinfo {author} {\bibfnamefont {M.}~\bibnamefont {Riallot}}, \bibinfo {author} {\bibfnamefont {B.}~\bibnamefont {Marini}}, \bibinfo {author} {\bibfnamefont {P.}~\bibnamefont {Gable}}, \bibinfo {author} {\bibfnamefont {Y.}~\bibnamefont {Date}}, \bibinfo {author} {\bibfnamefont {M.}~\bibnamefont {Sugiura}}, \bibinfo {author} {\bibfnamefont {Y.}~\bibnamefont {Elshayeb}}, \bibinfo {author} {\bibfnamefont {T.}~\bibnamefont {Elnady}}, \bibinfo {author} {\bibfnamefont {M.}~\bibnamefont {Ezzy}}, \bibinfo {author} {\bibfnamefont {E.}~\bibnamefont {Guerriero}}, \bibinfo {author} {\bibfnamefont {V.}~\bibnamefont {Steiger}}, \bibinfo {author} {\bibfnamefont {N.}~\bibnamefont {Serikoff}}, \bibinfo {author} {\bibfnamefont {J.-B.}\ \bibnamefont {Mouret}}, \bibinfo {author} {\bibfnamefont {B.}~\bibnamefont {Charlès}}, \bibinfo
  {author} {\bibfnamefont {H.}~\bibnamefont {Helal}},\ and\ \bibinfo {author} {\bibfnamefont {M.}~\bibnamefont {Tayoubi}},\ }\bibfield  {title} {\bibinfo {title} {Discovery of a big void in {Khufu}’s {Pyramid} by observation of cosmic-ray muons},\ }\href {https://doi.org/10.1038/nature24647} {\bibfield  {journal} {\bibinfo  {journal} {Nature}\ }\textbf {\bibinfo {volume} {552}},\ \bibinfo {pages} {386} (\bibinfo {year} {2017})}\BibitemShut {NoStop}%
\bibitem [{\citenamefont {Nishiyama}\ \emph {et~al.}(2017)\citenamefont {Nishiyama}, \citenamefont {Ariga}, \citenamefont {Ariga}, \citenamefont {Käser}, \citenamefont {Lechmann}, \citenamefont {Mair}, \citenamefont {Scampoli}, \citenamefont {Vladymyrov}, \citenamefont {Ereditato},\ and\ \citenamefont {Schlunegger}}]{nishiyama_first_2017}%
  \BibitemOpen
  \bibfield  {author} {\bibinfo {author} {\bibfnamefont {R.}~\bibnamefont {Nishiyama}}, \bibinfo {author} {\bibfnamefont {A.}~\bibnamefont {Ariga}}, \bibinfo {author} {\bibfnamefont {T.}~\bibnamefont {Ariga}}, \bibinfo {author} {\bibfnamefont {S.}~\bibnamefont {Käser}}, \bibinfo {author} {\bibfnamefont {A.}~\bibnamefont {Lechmann}}, \bibinfo {author} {\bibfnamefont {D.}~\bibnamefont {Mair}}, \bibinfo {author} {\bibfnamefont {P.}~\bibnamefont {Scampoli}}, \bibinfo {author} {\bibfnamefont {M.}~\bibnamefont {Vladymyrov}}, \bibinfo {author} {\bibfnamefont {A.}~\bibnamefont {Ereditato}},\ and\ \bibinfo {author} {\bibfnamefont {F.}~\bibnamefont {Schlunegger}},\ }\bibfield  {title} {\bibinfo {title} {First measurement of ice-bedrock interface of alpine glaciers by cosmic muon radiography},\ }\href {https://doi.org/10.1002/2017GL073599} {\bibfield  {journal} {\bibinfo  {journal} {Geophysical Research Letters}\ }\textbf {\bibinfo {volume} {44}},\ \bibinfo {pages} {6244} (\bibinfo {year} {2017})}\BibitemShut {NoStop}%
\bibitem [{\citenamefont {Fujii}\ \emph {et~al.}(2020)\citenamefont {Fujii}, \citenamefont {Hara}, \citenamefont {Hayashi}, \citenamefont {Kakuno}, \citenamefont {Kodama}, \citenamefont {Nagamine}, \citenamefont {Sato}, \citenamefont {Kim}, \citenamefont {Suzuki}, \citenamefont {Sumiyoshi}, \citenamefont {Takahashi}, \citenamefont {Takasaki}, \citenamefont {Tanaka},\ and\ \citenamefont {Yamashita}}]{fujii_investigation_2020}%
  \BibitemOpen
  \bibfield  {author} {\bibinfo {author} {\bibfnamefont {H.}~\bibnamefont {Fujii}}, \bibinfo {author} {\bibfnamefont {K.}~\bibnamefont {Hara}}, \bibinfo {author} {\bibfnamefont {K.}~\bibnamefont {Hayashi}}, \bibinfo {author} {\bibfnamefont {H.}~\bibnamefont {Kakuno}}, \bibinfo {author} {\bibfnamefont {H.}~\bibnamefont {Kodama}}, \bibinfo {author} {\bibfnamefont {K.}~\bibnamefont {Nagamine}}, \bibinfo {author} {\bibfnamefont {K.}~\bibnamefont {Sato}}, \bibinfo {author} {\bibfnamefont {S.-H.}\ \bibnamefont {Kim}}, \bibinfo {author} {\bibfnamefont {A.}~\bibnamefont {Suzuki}}, \bibinfo {author} {\bibfnamefont {T.}~\bibnamefont {Sumiyoshi}}, \bibinfo {author} {\bibfnamefont {K.}~\bibnamefont {Takahashi}}, \bibinfo {author} {\bibfnamefont {F.}~\bibnamefont {Takasaki}}, \bibinfo {author} {\bibfnamefont {S.}~\bibnamefont {Tanaka}},\ and\ \bibinfo {author} {\bibfnamefont {S.}~\bibnamefont {Yamashita}},\ }\bibfield  {title} {\bibinfo {title} {Investigation of the {Unit}-1 nuclear reactor of {Fukushima} {Daiichi} by
  cosmic muon radiography},\ }\href {https://doi.org/10.1093/ptep/ptaa027} {\bibfield  {journal} {\bibinfo  {journal} {Progress of Theoretical and Experimental Physics}\ }\textbf {\bibinfo {volume} {2020}},\ \bibinfo {pages} {043C02} (\bibinfo {year} {2020})}\BibitemShut {NoStop}%
\bibitem [{\citenamefont {Vanini}\ \emph {et~al.}(2018)\citenamefont {Vanini}, \citenamefont {Calvini}, \citenamefont {Checchia}, \citenamefont {Rigoni~Garola}, \citenamefont {Klinger}, \citenamefont {Zumerle}, \citenamefont {Bonomi}, \citenamefont {Donzella},\ and\ \citenamefont {Zenoni}}]{vanini_muography_2018}%
  \BibitemOpen
  \bibfield  {author} {\bibinfo {author} {\bibfnamefont {S.}~\bibnamefont {Vanini}}, \bibinfo {author} {\bibfnamefont {P.}~\bibnamefont {Calvini}}, \bibinfo {author} {\bibfnamefont {P.}~\bibnamefont {Checchia}}, \bibinfo {author} {\bibfnamefont {A.}~\bibnamefont {Rigoni~Garola}}, \bibinfo {author} {\bibfnamefont {J.}~\bibnamefont {Klinger}}, \bibinfo {author} {\bibfnamefont {G.}~\bibnamefont {Zumerle}}, \bibinfo {author} {\bibfnamefont {G.}~\bibnamefont {Bonomi}}, \bibinfo {author} {\bibfnamefont {A.}~\bibnamefont {Donzella}},\ and\ \bibinfo {author} {\bibfnamefont {A.}~\bibnamefont {Zenoni}},\ }\bibfield  {title} {\bibinfo {title} {Muography of different structures using muon scattering and absorption algorithms},\ }\href {https://doi.org/10.1098/rsta.2018.0051} {\bibfield  {journal} {\bibinfo  {journal} {Philosophical Transactions of the Royal Society A: Mathematical, Physical and Engineering Sciences}\ }\textbf {\bibinfo {volume} {377}},\ \bibinfo {pages} {20180051} (\bibinfo {year} {2018})}\BibitemShut
  {NoStop}%
\bibitem [{\citenamefont {Procureur}(2018)}]{procureur_muon_2018}%
  \BibitemOpen
  \bibfield  {author} {\bibinfo {author} {\bibfnamefont {S.}~\bibnamefont {Procureur}},\ }\bibfield  {title} {\bibinfo {title} {Muon imaging: {Principles}, technologies and applications},\ }\href {https://doi.org/10.1016/j.nima.2017.08.004} {\bibfield  {journal} {\bibinfo  {journal} {Nuclear Instruments and Methods in Physics Research Section A: Accelerators, Spectrometers, Detectors and Associated Equipment}\ }\bibinfo {series} {Radiation {Imaging} {Techniques} and {Applications}},\ \textbf {\bibinfo {volume} {878}},\ \bibinfo {pages} {169} (\bibinfo {year} {2018})}\BibitemShut {NoStop}%
\bibitem [{\citenamefont {Terzani}\ \emph {et~al.}(2025)\citenamefont {Terzani}, \citenamefont {Kisyov}, \citenamefont {Greenberg}, \citenamefont {Le~Pottier}, \citenamefont {Mironova}, \citenamefont {Picksley}, \citenamefont {Stackhouse}, \citenamefont {Tsai}, \citenamefont {Li}, \citenamefont {Rockafellow}, \citenamefont {Miao}, \citenamefont {Shrock}, \citenamefont {Heim}, \citenamefont {Garcia-Sciveres}, \citenamefont {Benedetti}, \citenamefont {Valentine}, \citenamefont {Milchberg}, \citenamefont {Nakamura}, \citenamefont {Gonsalves}, \citenamefont {van Tilborg}, \citenamefont {Schroeder}, \citenamefont {Esarey},\ and\ \citenamefont {Geddes}}]{terzani_measurement_2025}%
  \BibitemOpen
  \bibfield  {author} {\bibinfo {author} {\bibfnamefont {D.}~\bibnamefont {Terzani}}, \bibinfo {author} {\bibfnamefont {S.}~\bibnamefont {Kisyov}}, \bibinfo {author} {\bibfnamefont {S.}~\bibnamefont {Greenberg}}, \bibinfo {author} {\bibfnamefont {L.}~\bibnamefont {Le~Pottier}}, \bibinfo {author} {\bibfnamefont {M.}~\bibnamefont {Mironova}}, \bibinfo {author} {\bibfnamefont {A.}~\bibnamefont {Picksley}}, \bibinfo {author} {\bibfnamefont {J.}~\bibnamefont {Stackhouse}}, \bibinfo {author} {\bibfnamefont {H.-E.}\ \bibnamefont {Tsai}}, \bibinfo {author} {\bibfnamefont {R.}~\bibnamefont {Li}}, \bibinfo {author} {\bibfnamefont {E.}~\bibnamefont {Rockafellow}}, \bibinfo {author} {\bibfnamefont {B.}~\bibnamefont {Miao}}, \bibinfo {author} {\bibfnamefont {J.~E.}\ \bibnamefont {Shrock}}, \bibinfo {author} {\bibfnamefont {T.}~\bibnamefont {Heim}}, \bibinfo {author} {\bibfnamefont {M.}~\bibnamefont {Garcia-Sciveres}}, \bibinfo {author} {\bibfnamefont {C.}~\bibnamefont {Benedetti}}, \bibinfo {author} {\bibfnamefont
  {J.}~\bibnamefont {Valentine}}, \bibinfo {author} {\bibfnamefont {H.~M.}\ \bibnamefont {Milchberg}}, \bibinfo {author} {\bibfnamefont {K.}~\bibnamefont {Nakamura}}, \bibinfo {author} {\bibfnamefont {A.~J.}\ \bibnamefont {Gonsalves}}, \bibinfo {author} {\bibfnamefont {J.}~\bibnamefont {van Tilborg}}, \bibinfo {author} {\bibfnamefont {C.~B.}\ \bibnamefont {Schroeder}}, \bibinfo {author} {\bibfnamefont {E.}~\bibnamefont {Esarey}},\ and\ \bibinfo {author} {\bibfnamefont {C.~G.~R.}\ \bibnamefont {Geddes}},\ }\bibfield  {title} {\bibinfo {title} {Measurement of directional muon beams generated at the {Berkeley} {Lab} {Laser} {Accelerator}},\ }\href {https://doi.org/10.1103/kxjr-h7zs} {\bibfield  {journal} {\bibinfo  {journal} {Physical Review Accelerators and Beams}\ }\textbf {\bibinfo {volume} {28}},\ \bibinfo {pages} {103401} (\bibinfo {year} {2025})}\BibitemShut {NoStop}%
\bibitem [{\citenamefont {Zhang}\ \emph {et~al.}(2025)\citenamefont {Zhang}, \citenamefont {Deng}, \citenamefont {Ge}, \citenamefont {Wen}, \citenamefont {Cui}, \citenamefont {Feng}, \citenamefont {Wang}, \citenamefont {Wu}, \citenamefont {Pan}, \citenamefont {Liu}, \citenamefont {Deng}, \citenamefont {Zhang}, \citenamefont {Chen}, \citenamefont {Yan}, \citenamefont {Shan}, \citenamefont {Yuan}, \citenamefont {Tian}, \citenamefont {Qian}, \citenamefont {Zhu}, \citenamefont {Xu}, \citenamefont {Yu}, \citenamefont {Zhang}, \citenamefont {Yang}, \citenamefont {Zhou}, \citenamefont {Gu}, \citenamefont {Wang}, \citenamefont {Leng}, \citenamefont {Sun},\ and\ \citenamefont {Li}}]{zhang_proof--principle_2025}%
  \BibitemOpen
  \bibfield  {author} {\bibinfo {author} {\bibfnamefont {F.}~\bibnamefont {Zhang}}, \bibinfo {author} {\bibfnamefont {L.}~\bibnamefont {Deng}}, \bibinfo {author} {\bibfnamefont {Y.}~\bibnamefont {Ge}}, \bibinfo {author} {\bibfnamefont {J.}~\bibnamefont {Wen}}, \bibinfo {author} {\bibfnamefont {B.}~\bibnamefont {Cui}}, \bibinfo {author} {\bibfnamefont {K.}~\bibnamefont {Feng}}, \bibinfo {author} {\bibfnamefont {H.}~\bibnamefont {Wang}}, \bibinfo {author} {\bibfnamefont {C.}~\bibnamefont {Wu}}, \bibinfo {author} {\bibfnamefont {Z.}~\bibnamefont {Pan}}, \bibinfo {author} {\bibfnamefont {H.}~\bibnamefont {Liu}}, \bibinfo {author} {\bibfnamefont {Z.}~\bibnamefont {Deng}}, \bibinfo {author} {\bibfnamefont {Z.}~\bibnamefont {Zhang}}, \bibinfo {author} {\bibfnamefont {L.}~\bibnamefont {Chen}}, \bibinfo {author} {\bibfnamefont {D.}~\bibnamefont {Yan}}, \bibinfo {author} {\bibfnamefont {L.}~\bibnamefont {Shan}}, \bibinfo {author} {\bibfnamefont {Z.}~\bibnamefont {Yuan}}, \bibinfo {author} {\bibfnamefont
  {C.}~\bibnamefont {Tian}}, \bibinfo {author} {\bibfnamefont {J.}~\bibnamefont {Qian}}, \bibinfo {author} {\bibfnamefont {J.}~\bibnamefont {Zhu}}, \bibinfo {author} {\bibfnamefont {Y.}~\bibnamefont {Xu}}, \bibinfo {author} {\bibfnamefont {Y.}~\bibnamefont {Yu}}, \bibinfo {author} {\bibfnamefont {X.}~\bibnamefont {Zhang}}, \bibinfo {author} {\bibfnamefont {L.}~\bibnamefont {Yang}}, \bibinfo {author} {\bibfnamefont {W.}~\bibnamefont {Zhou}}, \bibinfo {author} {\bibfnamefont {Y.}~\bibnamefont {Gu}}, \bibinfo {author} {\bibfnamefont {W.}~\bibnamefont {Wang}}, \bibinfo {author} {\bibfnamefont {Y.}~\bibnamefont {Leng}}, \bibinfo {author} {\bibfnamefont {Z.}~\bibnamefont {Sun}},\ and\ \bibinfo {author} {\bibfnamefont {R.}~\bibnamefont {Li}},\ }\bibfield  {title} {\bibinfo {title} {Proof-of-principle demonstration of muon production with an ultrashort high-intensity laser},\ }\href {https://doi.org/10.1038/s41567-025-02872-2} {\bibfield  {journal} {\bibinfo  {journal} {Nature Physics}\ }\textbf {\bibinfo {volume}
  {21}},\ \bibinfo {pages} {1050} (\bibinfo {year} {2025})}\BibitemShut {NoStop}%
\bibitem [{\citenamefont {Chao}\ \emph {et~al.}(2013)\citenamefont {Chao}, \citenamefont {Mess},\ and\ \citenamefont {{others}}}]{chao_handbook_2013}%
  \BibitemOpen
  \bibfield  {author} {\bibinfo {author} {\bibfnamefont {A.~W.}\ \bibnamefont {Chao}}, \bibinfo {author} {\bibfnamefont {K.~H.}\ \bibnamefont {Mess}},\ and\ \bibinfo {author} {\bibnamefont {{others}}},\ }\href@noop {} {\emph {\bibinfo {title} {Handbook of accelerator physics and engineering}}}\ (\bibinfo  {publisher} {World scientific},\ \bibinfo {year} {2013})\BibitemShut {NoStop}%
\bibitem [{\citenamefont {Titov}\ \emph {et~al.}(2009)\citenamefont {Titov}, \citenamefont {Kämpfer},\ and\ \citenamefont {Takabe}}]{titov_dimuon_2009}%
  \BibitemOpen
  \bibfield  {author} {\bibinfo {author} {\bibfnamefont {A.~I.}\ \bibnamefont {Titov}}, \bibinfo {author} {\bibfnamefont {B.}~\bibnamefont {Kämpfer}},\ and\ \bibinfo {author} {\bibfnamefont {H.}~\bibnamefont {Takabe}},\ }\bibfield  {title} {\bibinfo {title} {Dimuon production by laser-wakefield accelerated electrons},\ }\href {https://doi.org/10.1103/PhysRevSTAB.12.111301} {\bibfield  {journal} {\bibinfo  {journal} {Physical Review Special Topics - Accelerators and Beams}\ }\textbf {\bibinfo {volume} {12}},\ \bibinfo {pages} {111301} (\bibinfo {year} {2009})}\BibitemShut {NoStop}%
\bibitem [{\citenamefont {Rao}\ \emph {et~al.}(2018)\citenamefont {Rao}, \citenamefont {Jeon}, \citenamefont {Kim},\ and\ \citenamefont {Nam}}]{rao_bright_2018}%
  \BibitemOpen
  \bibfield  {author} {\bibinfo {author} {\bibfnamefont {B.~S.}\ \bibnamefont {Rao}}, \bibinfo {author} {\bibfnamefont {J.~H.}\ \bibnamefont {Jeon}}, \bibinfo {author} {\bibfnamefont {H.~T.}\ \bibnamefont {Kim}},\ and\ \bibinfo {author} {\bibfnamefont {C.~H.}\ \bibnamefont {Nam}},\ }\bibfield  {title} {\bibinfo {title} {Bright muon source driven by {GeV} electron beams from a compact laser wakefield accelerator},\ }\href {https://doi.org/10.1088/1361-6587/aacdea} {\bibfield  {journal} {\bibinfo  {journal} {Plasma Physics and Controlled Fusion}\ }\textbf {\bibinfo {volume} {60}},\ \bibinfo {pages} {095002} (\bibinfo {year} {2018})}\BibitemShut {NoStop}%
\bibitem [{\citenamefont {Calvin}\ \emph {et~al.}(2023)\citenamefont {Calvin}, \citenamefont {Tomassini}, \citenamefont {Doria}, \citenamefont {Martello}, \citenamefont {Deas},\ and\ \citenamefont {Sarri}}]{calvin_laser-driven_2023}%
  \BibitemOpen
  \bibfield  {author} {\bibinfo {author} {\bibfnamefont {L.}~\bibnamefont {Calvin}}, \bibinfo {author} {\bibfnamefont {P.}~\bibnamefont {Tomassini}}, \bibinfo {author} {\bibfnamefont {D.}~\bibnamefont {Doria}}, \bibinfo {author} {\bibfnamefont {D.}~\bibnamefont {Martello}}, \bibinfo {author} {\bibfnamefont {R.~M.}\ \bibnamefont {Deas}},\ and\ \bibinfo {author} {\bibfnamefont {G.}~\bibnamefont {Sarri}},\ }\bibfield  {title} {\bibinfo {title} {Laser-driven muon production for material inspection and imaging},\ }\bibfield  {journal} {\bibinfo  {journal} {Frontiers in Physics}\ }\textbf {\bibinfo {volume} {11}},\ \href {https://doi.org/10.3389/fphy.2023.1177486} {10.3389/fphy.2023.1177486} (\bibinfo {year} {2023})\BibitemShut {NoStop}%
\bibitem [{\citenamefont {Calvin}\ \emph {et~al.}(2026)\citenamefont {Calvin}, \citenamefont {Gerstmayr}, \citenamefont {Arran}, \citenamefont {Tudor}, \citenamefont {Foster}, \citenamefont {Fleck}, \citenamefont {Bergmann}, \citenamefont {Doria}, \citenamefont {Kettle}, \citenamefont {Maguire}, \citenamefont {Malka}, \citenamefont {Manek}, \citenamefont {Mangles}, \citenamefont {McKenna}, \citenamefont {Mihai}, \citenamefont {Popa}, \citenamefont {Ridgers}, \citenamefont {Sarma}, \citenamefont {Smolyanskiy}, \citenamefont {Wilson}, \citenamefont {Deas},\ and\ \citenamefont {Sarri}}]{calvin_experimental_2026}%
  \BibitemOpen
  \bibfield  {author} {\bibinfo {author} {\bibfnamefont {L.}~\bibnamefont {Calvin}}, \bibinfo {author} {\bibfnamefont {E.}~\bibnamefont {Gerstmayr}}, \bibinfo {author} {\bibfnamefont {C.}~\bibnamefont {Arran}}, \bibinfo {author} {\bibfnamefont {L.}~\bibnamefont {Tudor}}, \bibinfo {author} {\bibfnamefont {T.}~\bibnamefont {Foster}}, \bibinfo {author} {\bibfnamefont {K.}~\bibnamefont {Fleck}}, \bibinfo {author} {\bibfnamefont {B.}~\bibnamefont {Bergmann}}, \bibinfo {author} {\bibfnamefont {D.}~\bibnamefont {Doria}}, \bibinfo {author} {\bibfnamefont {B.}~\bibnamefont {Kettle}}, \bibinfo {author} {\bibfnamefont {H.}~\bibnamefont {Maguire}}, \bibinfo {author} {\bibfnamefont {V.}~\bibnamefont {Malka}}, \bibinfo {author} {\bibfnamefont {P.}~\bibnamefont {Manek}}, \bibinfo {author} {\bibfnamefont {S.~P.~D.}\ \bibnamefont {Mangles}}, \bibinfo {author} {\bibfnamefont {P.}~\bibnamefont {McKenna}}, \bibinfo {author} {\bibfnamefont {R.~E.}\ \bibnamefont {Mihai}}, \bibinfo {author} {\bibfnamefont {S.}~\bibnamefont {Popa}},
  \bibinfo {author} {\bibfnamefont {C.}~\bibnamefont {Ridgers}}, \bibinfo {author} {\bibfnamefont {J.}~\bibnamefont {Sarma}}, \bibinfo {author} {\bibfnamefont {P.}~\bibnamefont {Smolyanskiy}}, \bibinfo {author} {\bibfnamefont {R.}~\bibnamefont {Wilson}}, \bibinfo {author} {\bibfnamefont {R.~M.}\ \bibnamefont {Deas}},\ and\ \bibinfo {author} {\bibfnamefont {G.}~\bibnamefont {Sarri}},\ }\bibfield  {title} {\bibinfo {title} {Experimental evidence of production of directional muons from a laser-wakefield accelerator},\ }\href {https://doi.org/10.1088/1361-6587/ae4d05} {\bibfield  {journal} {\bibinfo  {journal} {Plasma Physics and Controlled Fusion}\ }\textbf {\bibinfo {volume} {68}},\ \bibinfo {pages} {035015} (\bibinfo {year} {2026})}\BibitemShut {NoStop}%
\bibitem [{\citenamefont {Dreesen}\ \emph {et~al.}(2014)\citenamefont {Dreesen}, \citenamefont {Green}, \citenamefont {Browder}, \citenamefont {Wood}, \citenamefont {Schwellenbach}, \citenamefont {Ditmire}, \citenamefont {Tiwari},\ and\ \citenamefont {Wagner}}]{dreesen_detection_2014}%
  \BibitemOpen
  \bibfield  {author} {\bibinfo {author} {\bibfnamefont {W.}~\bibnamefont {Dreesen}}, \bibinfo {author} {\bibfnamefont {J.~A.}\ \bibnamefont {Green}}, \bibinfo {author} {\bibfnamefont {M.}~\bibnamefont {Browder}}, \bibinfo {author} {\bibfnamefont {J.}~\bibnamefont {Wood}}, \bibinfo {author} {\bibfnamefont {D.}~\bibnamefont {Schwellenbach}}, \bibinfo {author} {\bibfnamefont {T.}~\bibnamefont {Ditmire}}, \bibinfo {author} {\bibfnamefont {G.}~\bibnamefont {Tiwari}},\ and\ \bibinfo {author} {\bibfnamefont {C.}~\bibnamefont {Wagner}},\ }\bibfield  {title} {\bibinfo {title} {Detection of petawatt laser-induced muon source for rapid high-{Z} material detection},\ }in\ \href {https://doi.org/10.1109/NSSMIC.2014.7431088} {\emph {\bibinfo {booktitle} {2014 {IEEE} {Nuclear} {Science} {Symposium} and {Medical} {Imaging} {Conference} ({NSS}/{MIC})}}}\ (\bibinfo {year} {2014})\ pp.\ \bibinfo {pages} {1--6}\BibitemShut {NoStop}%
\bibitem [{\citenamefont {Kiani}\ \emph {et~al.}(2023)\citenamefont {Kiani}, \citenamefont {Zhou}, \citenamefont {Bahk}, \citenamefont {Bromage}, \citenamefont {Bruhwiler}, \citenamefont {Campbell}, \citenamefont {Chang}, \citenamefont {Chowdhury}, \citenamefont {Downer}, \citenamefont {Du}, \citenamefont {Esarey}, \citenamefont {Galvanauskas}, \citenamefont {Galvin}, \citenamefont {Häfner}, \citenamefont {Hoffmann}, \citenamefont {Joshi}, \citenamefont {Kanskar}, \citenamefont {Lu}, \citenamefont {Menoni}, \citenamefont {Messerly}, \citenamefont {Mirov}, \citenamefont {Palmer}, \citenamefont {Pogorelsky}, \citenamefont {Polyanskiy}, \citenamefont {Power}, \citenamefont {Reagan}, \citenamefont {Rocca}, \citenamefont {Rothenberg}, \citenamefont {Schmidt}, \citenamefont {Sistrunk}, \citenamefont {Spinka}, \citenamefont {Tochitsky}, \citenamefont {Vafaei-Najafabadi}, \citenamefont {Tilborg}, \citenamefont {Wilcox}, \citenamefont {Zuegel},\ and\ \citenamefont {Geddes}}]{kiani_high_2023}%
  \BibitemOpen
  \bibfield  {author} {\bibinfo {author} {\bibfnamefont {L.}~\bibnamefont {Kiani}}, \bibinfo {author} {\bibfnamefont {T.}~\bibnamefont {Zhou}}, \bibinfo {author} {\bibfnamefont {S.-W.}\ \bibnamefont {Bahk}}, \bibinfo {author} {\bibfnamefont {J.}~\bibnamefont {Bromage}}, \bibinfo {author} {\bibfnamefont {D.}~\bibnamefont {Bruhwiler}}, \bibinfo {author} {\bibfnamefont {E.~M.}\ \bibnamefont {Campbell}}, \bibinfo {author} {\bibfnamefont {Z.}~\bibnamefont {Chang}}, \bibinfo {author} {\bibfnamefont {E.}~\bibnamefont {Chowdhury}}, \bibinfo {author} {\bibfnamefont {M.}~\bibnamefont {Downer}}, \bibinfo {author} {\bibfnamefont {Q.}~\bibnamefont {Du}}, \bibinfo {author} {\bibfnamefont {E.}~\bibnamefont {Esarey}}, \bibinfo {author} {\bibfnamefont {A.}~\bibnamefont {Galvanauskas}}, \bibinfo {author} {\bibfnamefont {T.}~\bibnamefont {Galvin}}, \bibinfo {author} {\bibfnamefont {C.}~\bibnamefont {Häfner}}, \bibinfo {author} {\bibfnamefont {D.}~\bibnamefont {Hoffmann}}, \bibinfo {author} {\bibfnamefont {C.}~\bibnamefont
  {Joshi}}, \bibinfo {author} {\bibfnamefont {M.}~\bibnamefont {Kanskar}}, \bibinfo {author} {\bibfnamefont {W.}~\bibnamefont {Lu}}, \bibinfo {author} {\bibfnamefont {C.}~\bibnamefont {Menoni}}, \bibinfo {author} {\bibfnamefont {M.}~\bibnamefont {Messerly}}, \bibinfo {author} {\bibfnamefont {S.~B.}\ \bibnamefont {Mirov}}, \bibinfo {author} {\bibfnamefont {M.}~\bibnamefont {Palmer}}, \bibinfo {author} {\bibfnamefont {I.}~\bibnamefont {Pogorelsky}}, \bibinfo {author} {\bibfnamefont {M.}~\bibnamefont {Polyanskiy}}, \bibinfo {author} {\bibfnamefont {E.}~\bibnamefont {Power}}, \bibinfo {author} {\bibfnamefont {B.}~\bibnamefont {Reagan}}, \bibinfo {author} {\bibfnamefont {J.}~\bibnamefont {Rocca}}, \bibinfo {author} {\bibfnamefont {J.}~\bibnamefont {Rothenberg}}, \bibinfo {author} {\bibfnamefont {B.~E.}\ \bibnamefont {Schmidt}}, \bibinfo {author} {\bibfnamefont {E.}~\bibnamefont {Sistrunk}}, \bibinfo {author} {\bibfnamefont {T.}~\bibnamefont {Spinka}}, \bibinfo {author} {\bibfnamefont {S.}~\bibnamefont
  {Tochitsky}}, \bibinfo {author} {\bibfnamefont {N.}~\bibnamefont {Vafaei-Najafabadi}}, \bibinfo {author} {\bibfnamefont {J.~v.}\ \bibnamefont {Tilborg}}, \bibinfo {author} {\bibfnamefont {R.}~\bibnamefont {Wilcox}}, \bibinfo {author} {\bibfnamefont {J.}~\bibnamefont {Zuegel}},\ and\ \bibinfo {author} {\bibfnamefont {C.}~\bibnamefont {Geddes}},\ }\bibfield  {title} {\bibinfo {title} {High average power ultrafast laser technologies for driving future advanced accelerators},\ }\href {https://doi.org/10.1088/1748-0221/18/08/T08006} {\bibfield  {journal} {\bibinfo  {journal} {Journal of Instrumentation}\ }\textbf {\bibinfo {volume} {18}}\bibinfo  {number} { (08)},\ \bibinfo {pages} {T08006}}\BibitemShut {NoStop}%
\bibitem [{\citenamefont {Picksley}\ \emph {et~al.}(2024)\citenamefont {Picksley}, \citenamefont {Stackhouse}, \citenamefont {Benedetti}, \citenamefont {Nakamura}, \citenamefont {Tsai}, \citenamefont {Li}, \citenamefont {Miao}, \citenamefont {Shrock}, \citenamefont {Rockafellow}, \citenamefont {Milchberg}, \citenamefont {Schroeder}, \citenamefont {van Tilborg}, \citenamefont {Esarey}, \citenamefont {Geddes},\ and\ \citenamefont {Gonsalves}}]{picksley_matched_2024}%
  \BibitemOpen
\bibfield  {number} {  }\bibfield  {author} {\bibinfo {author} {\bibfnamefont {A.}~\bibnamefont {Picksley}}, \bibinfo {author} {\bibfnamefont {J.}~\bibnamefont {Stackhouse}}, \bibinfo {author} {\bibfnamefont {C.}~\bibnamefont {Benedetti}}, \bibinfo {author} {\bibfnamefont {K.}~\bibnamefont {Nakamura}}, \bibinfo {author} {\bibfnamefont {H.~E.}\ \bibnamefont {Tsai}}, \bibinfo {author} {\bibfnamefont {R.}~\bibnamefont {Li}}, \bibinfo {author} {\bibfnamefont {B.}~\bibnamefont {Miao}}, \bibinfo {author} {\bibfnamefont {J.~E.}\ \bibnamefont {Shrock}}, \bibinfo {author} {\bibfnamefont {E.}~\bibnamefont {Rockafellow}}, \bibinfo {author} {\bibfnamefont {H.~M.}\ \bibnamefont {Milchberg}}, \bibinfo {author} {\bibfnamefont {C.~B.}\ \bibnamefont {Schroeder}}, \bibinfo {author} {\bibfnamefont {J.}~\bibnamefont {van Tilborg}}, \bibinfo {author} {\bibfnamefont {E.}~\bibnamefont {Esarey}}, \bibinfo {author} {\bibfnamefont {C.~G.~R.}\ \bibnamefont {Geddes}},\ and\ \bibinfo {author} {\bibfnamefont {A.~J.}\ \bibnamefont
  {Gonsalves}},\ }\bibfield  {title} {\bibinfo {title} {Matched {Guiding} and {Controlled} {Injection} in {Dark}-{Current}-{Free}, 10-{GeV}-{Class}, {Channel}-{Guided} {Laser}-{Plasma} {Accelerators}},\ }\href {https://doi.org/10.1103/PhysRevLett.133.255001} {\bibfield  {journal} {\bibinfo  {journal} {Physical Review Letters}\ }\textbf {\bibinfo {volume} {133}},\ \bibinfo {pages} {255001} (\bibinfo {year} {2024})}\BibitemShut {NoStop}%
\bibitem [{\citenamefont {Shalloo}\ \emph {et~al.}(2018)\citenamefont {Shalloo}, \citenamefont {Arran}, \citenamefont {Corner}, \citenamefont {Holloway}, \citenamefont {Jonnerby}, \citenamefont {Walczak}, \citenamefont {Milchberg},\ and\ \citenamefont {Hooker}}]{shalloo_hydrodynamic_2018}%
  \BibitemOpen
  \bibfield  {author} {\bibinfo {author} {\bibfnamefont {R.~J.}\ \bibnamefont {Shalloo}}, \bibinfo {author} {\bibfnamefont {C.}~\bibnamefont {Arran}}, \bibinfo {author} {\bibfnamefont {L.}~\bibnamefont {Corner}}, \bibinfo {author} {\bibfnamefont {J.}~\bibnamefont {Holloway}}, \bibinfo {author} {\bibfnamefont {J.}~\bibnamefont {Jonnerby}}, \bibinfo {author} {\bibfnamefont {R.}~\bibnamefont {Walczak}}, \bibinfo {author} {\bibfnamefont {H.~M.}\ \bibnamefont {Milchberg}},\ and\ \bibinfo {author} {\bibfnamefont {S.~M.}\ \bibnamefont {Hooker}},\ }\bibfield  {title} {\bibinfo {title} {Hydrodynamic optical-field-ionized plasma channels},\ }\href {https://doi.org/10.1103/PhysRevE.97.053203} {\bibfield  {journal} {\bibinfo  {journal} {Physical Review E}\ }\textbf {\bibinfo {volume} {97}},\ \bibinfo {pages} {053203} (\bibinfo {year} {2018})}\BibitemShut {NoStop}%
\bibitem [{\citenamefont {Shalloo}\ \emph {et~al.}(2019)\citenamefont {Shalloo}, \citenamefont {Arran}, \citenamefont {Picksley}, \citenamefont {von Boetticher}, \citenamefont {Corner}, \citenamefont {Holloway}, \citenamefont {Hine}, \citenamefont {Jonnerby}, \citenamefont {Milchberg}, \citenamefont {Thornton}, \citenamefont {Walczak},\ and\ \citenamefont {Hooker}}]{shalloo_low-density_2019}%
  \BibitemOpen
  \bibfield  {author} {\bibinfo {author} {\bibfnamefont {R.}~\bibnamefont {Shalloo}}, \bibinfo {author} {\bibfnamefont {C.}~\bibnamefont {Arran}}, \bibinfo {author} {\bibfnamefont {A.}~\bibnamefont {Picksley}}, \bibinfo {author} {\bibfnamefont {A.}~\bibnamefont {von Boetticher}}, \bibinfo {author} {\bibfnamefont {L.}~\bibnamefont {Corner}}, \bibinfo {author} {\bibfnamefont {J.}~\bibnamefont {Holloway}}, \bibinfo {author} {\bibfnamefont {G.}~\bibnamefont {Hine}}, \bibinfo {author} {\bibfnamefont {J.}~\bibnamefont {Jonnerby}}, \bibinfo {author} {\bibfnamefont {H.}~\bibnamefont {Milchberg}}, \bibinfo {author} {\bibfnamefont {C.}~\bibnamefont {Thornton}}, \bibinfo {author} {\bibfnamefont {R.}~\bibnamefont {Walczak}},\ and\ \bibinfo {author} {\bibfnamefont {S.}~\bibnamefont {Hooker}},\ }\bibfield  {title} {\bibinfo {title} {Low-density hydrodynamic optical-field-ionized plasma channels generated with an axicon lens},\ }\href {https://doi.org/10.1103/PhysRevAccelBeams.22.041302} {\bibfield  {journal} {\bibinfo
  {journal} {Physical Review Accelerators and Beams}\ }\textbf {\bibinfo {volume} {22}},\ \bibinfo {pages} {041302} (\bibinfo {year} {2019})}\BibitemShut {NoStop}%
\bibitem [{\citenamefont {Smartsev}\ \emph {et~al.}(2019)\citenamefont {Smartsev}, \citenamefont {Caizergues}, \citenamefont {Oubrerie}, \citenamefont {Gautier}, \citenamefont {Goddet}, \citenamefont {Tafzi}, \citenamefont {Phuoc}, \citenamefont {Malka},\ and\ \citenamefont {Thaury}}]{smartsev_axiparabola_2019}%
  \BibitemOpen
  \bibfield  {author} {\bibinfo {author} {\bibfnamefont {S.}~\bibnamefont {Smartsev}}, \bibinfo {author} {\bibfnamefont {C.}~\bibnamefont {Caizergues}}, \bibinfo {author} {\bibfnamefont {K.}~\bibnamefont {Oubrerie}}, \bibinfo {author} {\bibfnamefont {J.}~\bibnamefont {Gautier}}, \bibinfo {author} {\bibfnamefont {J.-P.}\ \bibnamefont {Goddet}}, \bibinfo {author} {\bibfnamefont {A.}~\bibnamefont {Tafzi}}, \bibinfo {author} {\bibfnamefont {K.~T.}\ \bibnamefont {Phuoc}}, \bibinfo {author} {\bibfnamefont {V.}~\bibnamefont {Malka}},\ and\ \bibinfo {author} {\bibfnamefont {C.}~\bibnamefont {Thaury}},\ }\bibfield  {title} {\bibinfo {title} {Axiparabola: a long-focal-depth, high-resolution mirror for broadband high-intensity lasers},\ }\href {https://doi.org/10.1364/OL.44.003414} {\bibfield  {journal} {\bibinfo  {journal} {Optics Letters}\ }\textbf {\bibinfo {volume} {44}},\ \bibinfo {pages} {3414} (\bibinfo {year} {2019})}\BibitemShut {NoStop}%
\bibitem [{\citenamefont {Morozov}\ \emph {et~al.}(2018)\citenamefont {Morozov}, \citenamefont {Goltsov}, \citenamefont {Chen}, \citenamefont {Scully},\ and\ \citenamefont {Suckewer}}]{morozov_ionization_2018}%
  \BibitemOpen
  \bibfield  {author} {\bibinfo {author} {\bibfnamefont {A.}~\bibnamefont {Morozov}}, \bibinfo {author} {\bibfnamefont {A.}~\bibnamefont {Goltsov}}, \bibinfo {author} {\bibfnamefont {Q.}~\bibnamefont {Chen}}, \bibinfo {author} {\bibfnamefont {M.}~\bibnamefont {Scully}},\ and\ \bibinfo {author} {\bibfnamefont {S.}~\bibnamefont {Suckewer}},\ }\bibfield  {title} {\bibinfo {title} {Ionization assisted self-guiding of femtosecond laser pulses},\ }\href {https://doi.org/10.1063/1.5021795} {\bibfield  {journal} {\bibinfo  {journal} {Physics of Plasmas}\ }\textbf {\bibinfo {volume} {25}},\ \bibinfo {pages} {053110} (\bibinfo {year} {2018})}\BibitemShut {NoStop}%
\bibitem [{\citenamefont {Picksley}\ \emph {et~al.}(2020)\citenamefont {Picksley}, \citenamefont {Alejo}, \citenamefont {Shalloo}, \citenamefont {Arran}, \citenamefont {von Boetticher}, \citenamefont {Corner}, \citenamefont {Holloway}, \citenamefont {Jonnerby}, \citenamefont {Jakobsson}, \citenamefont {Thornton}, \citenamefont {Walczak},\ and\ \citenamefont {Hooker}}]{picksley_meter-scale_2020}%
  \BibitemOpen
  \bibfield  {author} {\bibinfo {author} {\bibfnamefont {A.}~\bibnamefont {Picksley}}, \bibinfo {author} {\bibfnamefont {A.}~\bibnamefont {Alejo}}, \bibinfo {author} {\bibfnamefont {R.~J.}\ \bibnamefont {Shalloo}}, \bibinfo {author} {\bibfnamefont {C.}~\bibnamefont {Arran}}, \bibinfo {author} {\bibfnamefont {A.}~\bibnamefont {von Boetticher}}, \bibinfo {author} {\bibfnamefont {L.}~\bibnamefont {Corner}}, \bibinfo {author} {\bibfnamefont {J.~A.}\ \bibnamefont {Holloway}}, \bibinfo {author} {\bibfnamefont {J.}~\bibnamefont {Jonnerby}}, \bibinfo {author} {\bibfnamefont {O.}~\bibnamefont {Jakobsson}}, \bibinfo {author} {\bibfnamefont {C.}~\bibnamefont {Thornton}}, \bibinfo {author} {\bibfnamefont {R.}~\bibnamefont {Walczak}},\ and\ \bibinfo {author} {\bibfnamefont {S.~M.}\ \bibnamefont {Hooker}},\ }\bibfield  {title} {\bibinfo {title} {Meter-scale conditioned hydrodynamic optical-field-ionized plasma channels},\ }\href {https://doi.org/10.1103/PhysRevE.102.053201} {\bibfield  {journal} {\bibinfo  {journal}
  {Physical Review E}\ }\textbf {\bibinfo {volume} {102}},\ \bibinfo {pages} {053201} (\bibinfo {year} {2020})}\BibitemShut {NoStop}%
\bibitem [{\citenamefont {Feder}\ \emph {et~al.}(2020)\citenamefont {Feder}, \citenamefont {Miao}, \citenamefont {Shrock}, \citenamefont {Goffin},\ and\ \citenamefont {Milchberg}}]{feder_self-waveguiding_2020}%
  \BibitemOpen
  \bibfield  {author} {\bibinfo {author} {\bibfnamefont {L.}~\bibnamefont {Feder}}, \bibinfo {author} {\bibfnamefont {B.}~\bibnamefont {Miao}}, \bibinfo {author} {\bibfnamefont {J.~E.}\ \bibnamefont {Shrock}}, \bibinfo {author} {\bibfnamefont {A.}~\bibnamefont {Goffin}},\ and\ \bibinfo {author} {\bibfnamefont {H.~M.}\ \bibnamefont {Milchberg}},\ }\bibfield  {title} {\bibinfo {title} {Self-waveguiding of relativistic laser pulses in neutral gas channels},\ }\href {https://doi.org/10.1103/PhysRevResearch.2.043173} {\bibfield  {journal} {\bibinfo  {journal} {Physical Review Research}\ }\textbf {\bibinfo {volume} {2}},\ \bibinfo {pages} {043173} (\bibinfo {year} {2020})}\BibitemShut {NoStop}%
\bibitem [{\citenamefont {Nakamura}\ \emph {et~al.}(2017)\citenamefont {Nakamura}, \citenamefont {Mao}, \citenamefont {Gonsalves}, \citenamefont {Vincenti}, \citenamefont {Mittelberger}, \citenamefont {Daniels}, \citenamefont {Magana}, \citenamefont {Toth},\ and\ \citenamefont {Leemans}}]{nakamura_diagnostics_2017}%
  \BibitemOpen
  \bibfield  {author} {\bibinfo {author} {\bibfnamefont {K.}~\bibnamefont {Nakamura}}, \bibinfo {author} {\bibfnamefont {H.-S.}\ \bibnamefont {Mao}}, \bibinfo {author} {\bibfnamefont {A.~J.}\ \bibnamefont {Gonsalves}}, \bibinfo {author} {\bibfnamefont {H.}~\bibnamefont {Vincenti}}, \bibinfo {author} {\bibfnamefont {D.~E.}\ \bibnamefont {Mittelberger}}, \bibinfo {author} {\bibfnamefont {J.}~\bibnamefont {Daniels}}, \bibinfo {author} {\bibfnamefont {A.}~\bibnamefont {Magana}}, \bibinfo {author} {\bibfnamefont {C.}~\bibnamefont {Toth}},\ and\ \bibinfo {author} {\bibfnamefont {W.~P.}\ \bibnamefont {Leemans}},\ }\bibfield  {title} {\bibinfo {title} {Diagnostics, {Control} and {Performance} {Parameters} for the {BELLA} {High} {Repetition} {Rate} {Petawatt} {Class} {Laser}},\ }\href {https://doi.org/10.1109/JQE.2017.2708601} {\bibfield  {journal} {\bibinfo  {journal} {IEEE Journal of Quantum Electronics}\ }\textbf {\bibinfo {volume} {53}},\ \bibinfo {pages} {1} (\bibinfo {year} {2017})},\ \bibinfo {note} {conference
  Name: IEEE Journal of Quantum Electronics}\BibitemShut {NoStop}%
\bibitem [{\citenamefont {Alimonti}\ \emph {et~al.}(2025)\citenamefont {Alimonti}, \citenamefont {Andreazza}, \citenamefont {Arteche}, \citenamefont {Barbero}, \citenamefont {Barrillon}, \citenamefont {Beccherle}, \citenamefont {Bonomelli}, \citenamefont {Bilei}, \citenamefont {Bialas}, \citenamefont {Bortoletto}, \citenamefont {Calderini}, \citenamefont {Caratelli}, \citenamefont {Cassese}, \citenamefont {Christiansen}, \citenamefont {Conti}, \citenamefont {Crescioli}, \citenamefont {Daas}, \citenamefont {Damenti}, \citenamefont {D'Auria}, \citenamefont {De~Canio}, \citenamefont {De~Robertis}, \citenamefont {Demaria}, \citenamefont {DeWitt}, \citenamefont {Dieter}, \citenamefont {Dimitrievska}, \citenamefont {Erdmann}, \citenamefont {Esposito}, \citenamefont {Exarchou}, \citenamefont {Fougeron}, \citenamefont {Gaioni}, \citenamefont {Garcia-Sciveres}, \citenamefont {Gnani}, \citenamefont {Gozalez~Renteria}, \citenamefont {Grippo}, \citenamefont {Guardino}, \citenamefont {Hamer}, \citenamefont {Heim},
  \citenamefont {Hemperek}, \citenamefont {Hinterkeuser}, \citenamefont {Huiberts}, \citenamefont {Jara~Casas}, \citenamefont {John}, \citenamefont {Kampkötter}, \citenamefont {Karagounis}, \citenamefont {Kazas}, \citenamefont {Khwaira}, \citenamefont {Kluit}, \citenamefont {Koukola}, \citenamefont {Krieger}, \citenamefont {Krüger}, \citenamefont {Lalic}, \citenamefont {Lauritzen}, \citenamefont {Licciulli}, \citenamefont {Liu}, \citenamefont {Loddo}, \citenamefont {Lopez~Morillo}, \citenamefont {Lounis}, \citenamefont {Luongo}, \citenamefont {Manghisoni}, \citenamefont {Marconi}, \citenamefont {Marquez~Lasso}, \citenamefont {Marzocca}, \citenamefont {Mauer}, \citenamefont {Mekkaoui}, \citenamefont {Meng}, \citenamefont {Menichelli}, \citenamefont {Menouni}, \citenamefont {Minuti}, \citenamefont {Mironova}, \citenamefont {Miryala}, \citenamefont {Missiroli}, \citenamefont {Monteil}, \citenamefont {Moustakas}, \citenamefont {Muñoz~Chavero}, \citenamefont {Neue}, \citenamefont {Orfanelli}, \citenamefont
  {Paccagnella}, \citenamefont {Pacher}, \citenamefont {Palla}, \citenamefont {Palomo~Pinto}, \citenamefont {Papadopoulou}, \citenamefont {Paterno}, \citenamefont {Petri}, \citenamefont {Placidi}, \citenamefont {Plackett}, \citenamefont {Pradas}, \citenamefont {Pulli}, \citenamefont {Raciti}, \citenamefont {Ratti}, \citenamefont {Re}, \citenamefont {Rehman}, \citenamefont {Rymaszewski}, \citenamefont {Sander}, \citenamefont {Solal}, \citenamefont {Standke}, \citenamefont {Stugu}, \citenamefont {Thompson}, \citenamefont {Traversi}, \citenamefont {Vogrig}, \citenamefont {Vogt}, \citenamefont {Wang}, \citenamefont {Yang}, \citenamefont {Zdenko},\ and\ \citenamefont {collaboration}}]{alimonti_rd53_2025}%
  \BibitemOpen
  \bibfield  {author} {\bibinfo {author} {\bibfnamefont {G.}~\bibnamefont {Alimonti}}, \bibinfo {author} {\bibfnamefont {A.}~\bibnamefont {Andreazza}}, \bibinfo {author} {\bibfnamefont {F.}~\bibnamefont {Arteche}}, \bibinfo {author} {\bibfnamefont {M.}~\bibnamefont {Barbero}}, \bibinfo {author} {\bibfnamefont {P.}~\bibnamefont {Barrillon}}, \bibinfo {author} {\bibfnamefont {R.}~\bibnamefont {Beccherle}}, \bibinfo {author} {\bibfnamefont {G.}~\bibnamefont {Bonomelli}}, \bibinfo {author} {\bibfnamefont {G.}~\bibnamefont {Bilei}}, \bibinfo {author} {\bibfnamefont {W.}~\bibnamefont {Bialas}}, \bibinfo {author} {\bibfnamefont {D.}~\bibnamefont {Bortoletto}}, \bibinfo {author} {\bibfnamefont {G.}~\bibnamefont {Calderini}}, \bibinfo {author} {\bibfnamefont {A.}~\bibnamefont {Caratelli}}, \bibinfo {author} {\bibfnamefont {A.}~\bibnamefont {Cassese}}, \bibinfo {author} {\bibfnamefont {J.}~\bibnamefont {Christiansen}}, \bibinfo {author} {\bibfnamefont {E.}~\bibnamefont {Conti}}, \bibinfo {author} {\bibfnamefont
  {F.}~\bibnamefont {Crescioli}}, \bibinfo {author} {\bibfnamefont {M.}~\bibnamefont {Daas}}, \bibinfo {author} {\bibfnamefont {L.}~\bibnamefont {Damenti}}, \bibinfo {author} {\bibfnamefont {S.}~\bibnamefont {D'Auria}}, \bibinfo {author} {\bibfnamefont {F.}~\bibnamefont {De~Canio}}, \bibinfo {author} {\bibfnamefont {G.}~\bibnamefont {De~Robertis}}, \bibinfo {author} {\bibfnamefont {N.}~\bibnamefont {Demaria}}, \bibinfo {author} {\bibfnamefont {J.}~\bibnamefont {DeWitt}}, \bibinfo {author} {\bibfnamefont {Y.}~\bibnamefont {Dieter}}, \bibinfo {author} {\bibfnamefont {A.}~\bibnamefont {Dimitrievska}}, \bibinfo {author} {\bibfnamefont {W.}~\bibnamefont {Erdmann}}, \bibinfo {author} {\bibfnamefont {S.}~\bibnamefont {Esposito}}, \bibinfo {author} {\bibfnamefont {D.}~\bibnamefont {Exarchou}}, \bibinfo {author} {\bibfnamefont {D.}~\bibnamefont {Fougeron}}, \bibinfo {author} {\bibfnamefont {L.}~\bibnamefont {Gaioni}}, \bibinfo {author} {\bibfnamefont {M.}~\bibnamefont {Garcia-Sciveres}}, \bibinfo {author}
  {\bibfnamefont {D.}~\bibnamefont {Gnani}}, \bibinfo {author} {\bibfnamefont {C.}~\bibnamefont {Gozalez~Renteria}}, \bibinfo {author} {\bibfnamefont {M.}~\bibnamefont {Grippo}}, \bibinfo {author} {\bibfnamefont {A.}~\bibnamefont {Guardino}}, \bibinfo {author} {\bibfnamefont {M.}~\bibnamefont {Hamer}}, \bibinfo {author} {\bibfnamefont {T.}~\bibnamefont {Heim}}, \bibinfo {author} {\bibfnamefont {T.}~\bibnamefont {Hemperek}}, \bibinfo {author} {\bibfnamefont {F.}~\bibnamefont {Hinterkeuser}}, \bibinfo {author} {\bibfnamefont {S.}~\bibnamefont {Huiberts}}, \bibinfo {author} {\bibfnamefont {L.}~\bibnamefont {Jara~Casas}}, \bibinfo {author} {\bibfnamefont {J.}~\bibnamefont {John}}, \bibinfo {author} {\bibfnamefont {J.}~\bibnamefont {Kampkötter}}, \bibinfo {author} {\bibfnamefont {M.}~\bibnamefont {Karagounis}}, \bibinfo {author} {\bibfnamefont {I.}~\bibnamefont {Kazas}}, \bibinfo {author} {\bibfnamefont {Y.}~\bibnamefont {Khwaira}}, \bibinfo {author} {\bibfnamefont {R.}~\bibnamefont {Kluit}}, \bibinfo {author}
  {\bibfnamefont {D.}~\bibnamefont {Koukola}}, \bibinfo {author} {\bibfnamefont {A.}~\bibnamefont {Krieger}}, \bibinfo {author} {\bibfnamefont {H.}~\bibnamefont {Krüger}}, \bibinfo {author} {\bibfnamefont {J.}~\bibnamefont {Lalic}}, \bibinfo {author} {\bibfnamefont {M.}~\bibnamefont {Lauritzen}}, \bibinfo {author} {\bibfnamefont {F.}~\bibnamefont {Licciulli}}, \bibinfo {author} {\bibfnamefont {P.}~\bibnamefont {Liu}}, \bibinfo {author} {\bibfnamefont {F.}~\bibnamefont {Loddo}}, \bibinfo {author} {\bibfnamefont {E.}~\bibnamefont {Lopez~Morillo}}, \bibinfo {author} {\bibfnamefont {A.}~\bibnamefont {Lounis}}, \bibinfo {author} {\bibfnamefont {F.}~\bibnamefont {Luongo}}, \bibinfo {author} {\bibfnamefont {M.}~\bibnamefont {Manghisoni}}, \bibinfo {author} {\bibfnamefont {S.}~\bibnamefont {Marconi}}, \bibinfo {author} {\bibfnamefont {F.}~\bibnamefont {Marquez~Lasso}}, \bibinfo {author} {\bibfnamefont {C.}~\bibnamefont {Marzocca}}, \bibinfo {author} {\bibfnamefont {K.}~\bibnamefont {Mauer}}, \bibinfo {author}
  {\bibfnamefont {A.}~\bibnamefont {Mekkaoui}}, \bibinfo {author} {\bibfnamefont {L.}~\bibnamefont {Meng}}, \bibinfo {author} {\bibfnamefont {M.}~\bibnamefont {Menichelli}}, \bibinfo {author} {\bibfnamefont {M.}~\bibnamefont {Menouni}}, \bibinfo {author} {\bibfnamefont {M.}~\bibnamefont {Minuti}}, \bibinfo {author} {\bibfnamefont {M.}~\bibnamefont {Mironova}}, \bibinfo {author} {\bibfnamefont {S.}~\bibnamefont {Miryala}}, \bibinfo {author} {\bibfnamefont {M.}~\bibnamefont {Missiroli}}, \bibinfo {author} {\bibfnamefont {E.}~\bibnamefont {Monteil}}, \bibinfo {author} {\bibfnamefont {K.}~\bibnamefont {Moustakas}}, \bibinfo {author} {\bibfnamefont {F.}~\bibnamefont {Muñoz~Chavero}}, \bibinfo {author} {\bibfnamefont {G.}~\bibnamefont {Neue}}, \bibinfo {author} {\bibfnamefont {S.}~\bibnamefont {Orfanelli}}, \bibinfo {author} {\bibfnamefont {A.}~\bibnamefont {Paccagnella}}, \bibinfo {author} {\bibfnamefont {L.}~\bibnamefont {Pacher}}, \bibinfo {author} {\bibfnamefont {F.}~\bibnamefont {Palla}}, \bibinfo {author}
  {\bibfnamefont {F.}~\bibnamefont {Palomo~Pinto}}, \bibinfo {author} {\bibfnamefont {A.}~\bibnamefont {Papadopoulou}}, \bibinfo {author} {\bibfnamefont {A.}~\bibnamefont {Paterno}}, \bibinfo {author} {\bibfnamefont {A.}~\bibnamefont {Petri}}, \bibinfo {author} {\bibfnamefont {P.}~\bibnamefont {Placidi}}, \bibinfo {author} {\bibfnamefont {R.}~\bibnamefont {Plackett}}, \bibinfo {author} {\bibfnamefont {A.}~\bibnamefont {Pradas}}, \bibinfo {author} {\bibfnamefont {A.}~\bibnamefont {Pulli}}, \bibinfo {author} {\bibfnamefont {B.}~\bibnamefont {Raciti}}, \bibinfo {author} {\bibfnamefont {L.}~\bibnamefont {Ratti}}, \bibinfo {author} {\bibfnamefont {V.}~\bibnamefont {Re}}, \bibinfo {author} {\bibfnamefont {A.}~\bibnamefont {Rehman}}, \bibinfo {author} {\bibfnamefont {P.}~\bibnamefont {Rymaszewski}}, \bibinfo {author} {\bibfnamefont {P.}~\bibnamefont {Sander}}, \bibinfo {author} {\bibfnamefont {M.}~\bibnamefont {Solal}}, \bibinfo {author} {\bibfnamefont {M.}~\bibnamefont {Standke}}, \bibinfo {author} {\bibfnamefont
  {B.}~\bibnamefont {Stugu}}, \bibinfo {author} {\bibfnamefont {E.}~\bibnamefont {Thompson}}, \bibinfo {author} {\bibfnamefont {G.}~\bibnamefont {Traversi}}, \bibinfo {author} {\bibfnamefont {D.}~\bibnamefont {Vogrig}}, \bibinfo {author} {\bibfnamefont {M.}~\bibnamefont {Vogt}}, \bibinfo {author} {\bibfnamefont {T.}~\bibnamefont {Wang}}, \bibinfo {author} {\bibfnamefont {H.}~\bibnamefont {Yang}}, \bibinfo {author} {\bibfnamefont {J.}~\bibnamefont {Zdenko}},\ and\ \bibinfo {author} {\bibfnamefont {T.~R.}\ \bibnamefont {collaboration}},\ }\bibfield  {title} {\bibinfo {title} {{RD53} pixel readout integrated circuits for {ATLAS} and {CMS} {HL}-{LHC} upgrades},\ }\href {https://doi.org/10.1088/1748-0221/20/03/P03024} {\bibfield  {journal} {\bibinfo  {journal} {Journal of Instrumentation}\ }\textbf {\bibinfo {volume} {20}}\bibinfo  {number} { (03)},\ \bibinfo {pages} {P03024}}\BibitemShut {NoStop}%
\bibitem [{\citenamefont {Samy}\ \emph {et~al.}(2025)\citenamefont {Samy}, \citenamefont {Calderini}, \citenamefont {Carcone}, \citenamefont {Carlotto}, \citenamefont {Chabrillat}, \citenamefont {Gemme}, \citenamefont {Grigorev}, \citenamefont {Heim}, \citenamefont {Meng}, \citenamefont {Miranova}, \citenamefont {Ravera}, \citenamefont {Ressegotti}, \citenamefont {Rummler}, \citenamefont {Skaf},\ and\ \citenamefont {Krause}}]{samy_recent_2025}%
  \BibitemOpen
\bibfield  {number} {  }\bibfield  {author} {\bibinfo {author} {\bibfnamefont {M.~A.~A.}\ \bibnamefont {Samy}}, \bibinfo {author} {\bibfnamefont {G.}~\bibnamefont {Calderini}}, \bibinfo {author} {\bibfnamefont {T.}~\bibnamefont {Carcone}}, \bibinfo {author} {\bibfnamefont {J.}~\bibnamefont {Carlotto}}, \bibinfo {author} {\bibfnamefont {P.}~\bibnamefont {Chabrillat}}, \bibinfo {author} {\bibfnamefont {C.}~\bibnamefont {Gemme}}, \bibinfo {author} {\bibfnamefont {A.}~\bibnamefont {Grigorev}}, \bibinfo {author} {\bibfnamefont {T.}~\bibnamefont {Heim}}, \bibinfo {author} {\bibfnamefont {L.}~\bibnamefont {Meng}}, \bibinfo {author} {\bibfnamefont {M.}~\bibnamefont {Miranova}}, \bibinfo {author} {\bibfnamefont {S.}~\bibnamefont {Ravera}}, \bibinfo {author} {\bibfnamefont {M.}~\bibnamefont {Ressegotti}}, \bibinfo {author} {\bibfnamefont {A.}~\bibnamefont {Rummler}}, \bibinfo {author} {\bibfnamefont {A.}~\bibnamefont {Skaf}},\ and\ \bibinfo {author} {\bibfnamefont {C.}~\bibnamefont {Krause}},\ }\bibfield  {title}
  {\bibinfo {title} {Recent test beam results of {ATLAS} {ITk} pixel modules},\ }\href {https://doi.org/10.1088/1748-0221/20/06/C06025} {\bibfield  {journal} {\bibinfo  {journal} {Journal of Instrumentation}\ }\textbf {\bibinfo {volume} {20}}\bibinfo  {number} { (06)},\ \bibinfo {pages} {C06025}}\BibitemShut {NoStop}%
\bibitem [{\citenamefont {Dannheim}\ \emph {et~al.}(2021)\citenamefont {Dannheim}, \citenamefont {Dort}, \citenamefont {Huth}, \citenamefont {Hynds}, \citenamefont {Kremastiotis}, \citenamefont {Kröger}, \citenamefont {Munker}, \citenamefont {Pitters}, \citenamefont {Schütze}, \citenamefont {Spannagel}, \citenamefont {Vanat},\ and\ \citenamefont {Williams}}]{dannheim_corryvreckan_2021}%
  \BibitemOpen
\bibfield  {number} {  }\bibfield  {author} {\bibinfo {author} {\bibfnamefont {D.}~\bibnamefont {Dannheim}}, \bibinfo {author} {\bibfnamefont {K.}~\bibnamefont {Dort}}, \bibinfo {author} {\bibfnamefont {L.}~\bibnamefont {Huth}}, \bibinfo {author} {\bibfnamefont {D.}~\bibnamefont {Hynds}}, \bibinfo {author} {\bibfnamefont {I.}~\bibnamefont {Kremastiotis}}, \bibinfo {author} {\bibfnamefont {J.}~\bibnamefont {Kröger}}, \bibinfo {author} {\bibfnamefont {M.}~\bibnamefont {Munker}}, \bibinfo {author} {\bibfnamefont {F.}~\bibnamefont {Pitters}}, \bibinfo {author} {\bibfnamefont {P.}~\bibnamefont {Schütze}}, \bibinfo {author} {\bibfnamefont {S.}~\bibnamefont {Spannagel}}, \bibinfo {author} {\bibfnamefont {T.}~\bibnamefont {Vanat}},\ and\ \bibinfo {author} {\bibfnamefont {M.}~\bibnamefont {Williams}},\ }\bibfield  {title} {\bibinfo {title} {Corryvreckan: a modular {4D} track reconstruction and analysis software for test beam data},\ }\href {https://doi.org/10.1088/1748-0221/16/03/P03008} {\bibfield  {journal}
  {\bibinfo  {journal} {Journal of Instrumentation}\ }\textbf {\bibinfo {volume} {16}}\bibinfo  {number} { (03)},\ \bibinfo {pages} {P03008}}\BibitemShut {NoStop}%
\end{thebibliography}%
\end{document}